\documentclass[acmsmall]{acmart}

\usepackage{multirow}
\usepackage{algorithm}
\usepackage{algpseudocode}
\usepackage{bm}

\AtBeginDocument{%
  }

\setcopyright{acmlicensed}
\copyrightyear{2024}
\acmYear{2024}
\acmDOI{XXXXXXX.XXXXXXX}

\acmJournal{TOIS}

\begin{document}

\title{Adversarial Item Promotion on Visually-Aware Recommender Systems by Guided Diffusion}

\author{Lijian Chen}
\authornote{Equal contributions.}
\affiliation{%
  \institution{The University of Queensland}
  \city{Brisbane}
  \state{QLD}
  \country{Australia}
}
\email{uqlche22@uq.edu.au}

\author{Wei Yuan}
\authornotemark[1]
\affiliation{%
  \institution{The University of Queensland}
  \city{Brisbane}
  \state{QLD}
  \country{Australia}
}
\email{w.yuan@uq.edu.au}

\author{Tong Chen}
\affiliation{%
  \institution{The University of Queensland}
  \city{Brisbane}
  \state{QLD}
  \country{Australia}
}
\email{tong.chen@uq.edu.au}

\author{Guanhua Ye}
\affiliation{%
 \institution{Deep Neural Computing Company Limited}
 \city{Shenzhen}
 \country{China}}
\email{rex.ye@dncc.tech}

\author{Quoc Viet Hung Nguyen}
\affiliation{%
  \institution{Griffith University}
  \city{Gold Coast}
  \state{QLD}
  \country{Australia}
}
\email{henry.nguyen@griffith.edu.au}

\author{Hongzhi Yin}\authornote{Corresponding author.}
\affiliation{%
  \institution{The University of Queensland}
  \city{Brisbane}
  \state{QLD}
  \country{Australia}
}
\email{db.hongzhi@gmail.com}

\renewcommand{\shortauthors}{Chen et al.}

\begin{abstract}
Visually-aware recommender systems have found widespread applications in domains where visual elements significantly contribute to the inference of users' potential preferences. While the incorporation of visual information holds the promise of enhancing recommendation accuracy and alleviating the cold-start problem, it is essential to point out that the inclusion of item images may introduce substantial security challenges. Some existing works have shown that the item provider can manipulate item exposure rates to its advantage by constructing adversarial images. However, these works cannot reveal the real vulnerability of visually-aware recommender systems because (1) the generated adversarial images are markedly distorted, rendering them easily detected by human observers; (2) the effectiveness of these attacks is inconsistent and even ineffective in some scenarios or datasets. To shed light on the real vulnerabilities of visually-aware recommender systems when confronted with adversarial images, this paper introduces a novel attack method, IPDGI (Item Promotion by Diffusion Generated Image). Specifically, IPDGI employs a guided diffusion model to generate adversarial samples designed to promote the exposure rates of target items (e.g., long-tail items). Taking advantage of accurately modeling benign images' distribution by diffusion models, the generated adversarial images have high fidelity with original images, ensuring the stealth of our IPDGI. To demonstrate the effectiveness of our proposed methods, we conduct extensive experiments on two commonly used e-commerce recommendation datasets (Amazon Beauty and Amazon Baby) with several typical visually-aware recommender systems. The experimental results show that our attack method significantly improves both the performance of promoting the long-tailed (i.e., unpopular) items and the quality of generated adversarial images.
\end{abstract}




\maketitle

\section{Introduction}
With the exponential growth of data, recommender systems have become nearly indispensable across various industry sectors due to their ability to provide personalized suggestions~\cite{li2021lightweight,yin2015joint,yuan2023manipulating}. Traditionally, recommender systems predict users' preferences by learning user and item latent features from extensive collaborative data, such as user-item interactions~\cite{chen2018tada, yin2024device}. While these traditional recommender systems have achieved notable success, they often fall short in certain domains where users' preferences and decisions are strongly influenced by visual factors, such as fashion, food, and micro-video recommendations~\cite{cheng2023image}. Furthermore, given the persistent reality of data sparsity within collaborative datasets, traditional recommender systems struggle with cold-start problems~\cite{schein2002methods, zheng2023automl}. To address these two challenges, researchers have incorporated item visual information to assist systems in making recommendations, giving rise to visually-aware recommender systems~\cite{he2016vbpr,kang2017visually,tang2019adversarial}.

While harnessing visual features offers numerous advantages, incorporating these features may also introduce vulnerabilities in visually-aware recommender systems. Numerous existing works~\cite{long2022survey,nguyen2022survey,zeng2019adversarial} in the field of computer vision have demonstrated that by constructing adversarial images, even state-of-the-art deep neural network models can be disrupted by adversaries. These adversarial images look like normal images with imperceptible perturbations that are carefully crafted by adversaries according to specific objectives. As it is challenging to distinguish adversarial images from normal ones~\cite{elsayed2018adversarial}, such kind of adversarial attacks pose serious threats to the application of computer vision models. In recommender systems, the number of items can be more than the million level~\cite{yin2014lcars}, and item images are usually provided by external parties (e.g., item merchants) on social media platforms and E-commerce platforms. This setting leaves a backdoor for untrusted image providers to upload poisoned images to achieve certain adversarial goals, such as promoting the target item's ranking with financial incentives. In light of this, it is necessary to validate the threats of adversarial images in visually-aware recommender systems.

\begin{figure*}
    \centering
    \includegraphics[scale=0.18]{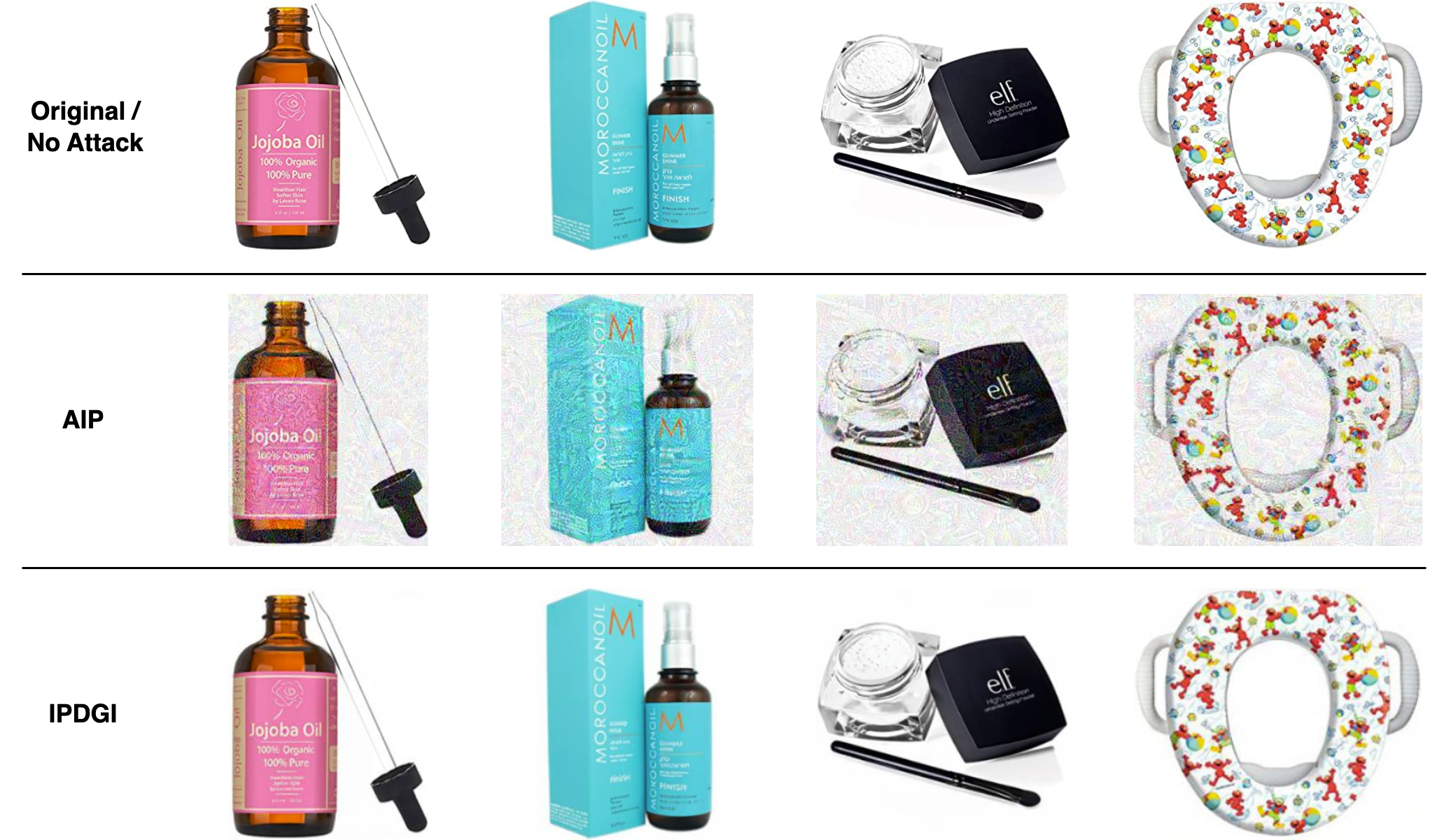}
    \caption{Image Comparisons of Original, AIP Attack (Baseline), and IPDGI Attack (Ours)}
    \label{fig_image_comparisons}
\end{figure*}

Existing attacks in visually-aware recommender systems generally can be categorized into two types: classifier-targeted attacks and ranker-targeted attacks. The classifier-targeted attack~\cite{di2020taamr} aims to change the prediction of item categories, which cannot directly change items' ranking. In contrast, the ranker-targeted attack uses adversarial samples to directly manipulate the top-K recommendation ranker. Liu et al.~\cite{liu2021adversarial} is the first work to investigate how to deceive recommender systems via perturbed visual information. However, their approach exhibits loose constraints regarding the scale of noise added to adversarial images, rendering their attacks impractical as their generated images are easily detectable by users. Figure~\ref{fig_image_comparisons} illustrates the adversarial images generated by their proposed AIP attack. Furthermore, the efficacy of these attacks tends to be highly unstable.~\cite{cohen2021black} explores a black-box setting of adversarial image attack. Nevertheless, this approach heavily relies on the assumption that each target user has a surrogate user, a condition that may not always hold in many practical scenarios since the untrusted third party (e.g., item merchants) has less chance to know the target user's whole interaction behavior. As a result, all existing adversarial image attacks in visually-aware recommender systems cannot reveal the real threats as they are ineffective with realistic conditions.

The primary objective of this work is to develop an adversarial attack~\cite{nguyen2024manipulating} to disclose the real vulnerability of visually-aware recommender systems, highlighting the security concerns of using images provided by third parties. To achieve that, the adversarial image attack should meet the following two requirements. Firstly, the attack should be both effective and inconspicuous. In other words, the generated adversarial image should closely resemble the original images while successfully misleading the recommender system. Secondly, the underlying assumptions of the attack methodology should align with real-world conditions and constraints.

Diffusion models have garnered remarkable success in the field of image generation~\cite{dhariwal2021diffusion}. Drawing inspiration from their impressive capacity to model real data distributions, we endeavor to exploit diffusion models for the purpose of generating adversarial images. Since the generated adversarial images are still from the normal image distribution, these poisoned images would be imperceptible, ensuring the stealthiness of our attack. Nevertheless, constructing a diffusion model-based attack tailored for visually-aware recommender systems presents at least two challenges. The first challenge is how to keep the consistency of generated adversarial images with original images. Due to the randomness of the diffusion model, the content of images generated by the general diffusion model is random, leading to a wide variation in the generated adversarial images that noticeably differ from the original images. Besides, how to incorporate the adversarial goal in the general diffusion model, i.e., ensure the effectiveness of the generated adversarial images from the diffusion model, is also non-trivial.

In this paper, we propose a novel adversarial attack method, namely \textit{Item Promotion by Diffusion Generated Images (IPDGI)}. To address the generation randomness of the diffusion model, we introduce a conditional constraint into the reverse process. This constraint ensures that the generated adversarial images are as similar as possible to the original image in terms of data distribution and visual appearance. To enhance the effectiveness of adversarial images generated by the diffusion model in the context of visually-aware recommender systems, we have devised a novel mechanism for perturbation generation, as shown in Figure~\ref{fig_perturbation_gen}. In essence, the underlying concept is to perturb the target item image to align its visual features with those of popular items. The perturbation generation mechanism involves several key steps. Initially, a clustering model is employed to identify the cluster to which the target item image belongs. Subsequently, we select the image of the most popular item within that cluster as a reference image. Following this, we iteratively optimize the perturbation to align the feature vector of the target item image with that of the reference image. It is crucial to note that, as the reference image is chosen from the same cluster, it is semantically close to our target image, and a slight perturbation is adequate to align the feature vectors of our target image. Nevertheless, the issue of severe distortion might still exist in the adversarial image if we directly apply the perturbation to the original image. To overcome this challenge, we integrate the optimized perturbation into the diffusion model's general Gaussian noise, as shown in Figure~\ref{fig_ipdgi}. By doing this, the corrupted image generated by the forward process of the diffusion model contains the perturbation. Later on, during the reverse process, as the diffusion model was trained on normal images, the model tends to denoise the image to the normal/clean image domain while preserving the perturbation. Ultimately, the adversarial images generated by the IPDGI are capable of deceiving the top-K ranker for item promotion~\cite{zheng2024poisoning, yuan2024robust} while simultaneously maintaining a high similarity to the original images, as shown in Figure~\ref{fig_image_comparisons}.

To validate the effectiveness of our proposed IPDGI, we conduct extensive experiments on two widely used recommendation datasets with three visually-aware recommender systems. The experimental results demonstrate the effectiveness of our method, IPDGI, in promoting items across all experimental datasets and visually-aware recommender systems. Our method outperforms existing ranker-targeted attacks. Furthermore, the experimental results indicate that the side effects caused by IPDGI are minimal, i.e., the original performance of the recommender system undergoes no significant changes under the worst-case attack scenario. Lastly, the image quality of the adversarial images generated by IPDGI surpasses that of the baseline attack, showcasing a noticeable improvement.

To sum up, the main contributions of this paper are as follows:
\begin{itemize}
    \item This is the first work to employ the diffusion model for generating adversarial samples against the visually-aware recommender systems for item promotion.
    \item We reveal the real vulnerability of the visually-aware recommender system with respect to the utilization of visual features.
    \item We evaluate our method, IPDGI, across three representative visually-aware recommender systems on two real-world datasets to assess the effectiveness and stealthiness of the attack.
\end{itemize}

The following sections of the paper are organized as follows: In Section~\ref{related_work}, we conduct a review of related work. Following that, Section~\ref{preliminaries} discusses the preliminaries of this research, encompassing the base visually-aware recommender systems and the adversarial approaches of IPDGI. Section~\ref{our_approach} is dedicated to presenting the technical details of our method, IPDGI. Transitioning to Section~\ref{experiments}, we provide comprehensive details on the experiments and results. Section~\ref{sec_defense} discusses the potential defense methods. Finally, in Section~\ref{conclusion}, we draw conclusions from our work.

\section{Related Work}\label{related_work}
\subsection{Visually-Aware Recommender Systems}
Visually-aware recommender systems are those that incorporate visual information into the recommendation ranking mechanism or the prediction of the user's preference. Initially, before the deep learning era, most of the works that adopted visual information relied on image retrieval for the recommendation tasks. Kalantidis et al.~\cite{kalantidis2013getting} propose an approach that commences with the segmentation of a query image, followed by the retrieval of visually similar items within each of the predicted classes. This work integrates semantic information derived from the images to enhance retrieval performance. Following this, Jagadeesh et al.~\cite{jagadeesh2014large} emphasized the pivotal role of semantic information in the retrieval process, highlighting its significance in refining and enhancing the overall efficiency of the retrieval procedure. In this work, they curated the extensive Fashion-136K dataset, enriched with detailed annotations, and introduced multiple retrieval-based methodologies to recommend matching items corresponding to a given query image.

With the advancements in Convolutional Neural Networks (CNNs)~\cite{he2016deep, simonyan2014very} and the developments of deep learning-based recommender systems~\cite{he2017neural, qiu2019rethinking, qiu2020gag,qu2021imgagn}, numerous studies have concentrated on more intricate modeling that integrates visual features into user-item interactions. IBR~\cite{mcauley2015image} suggests complementary items by analyzing the styles inherent in the visual features of each item, taking into consideration human perceptions of similarity. Later on, several works further incorporate visual information into Collaborative Filtering (CF)-based recommender models to exploit the latent factors of users and items along with visual features simultaneously. Examples include VBPR~\cite{he2016vbpr} and Fashion DNA~\cite{bracher2016fashion}. Notably, VBPR is the pioneer in integrating the pre-extracted CNN visual features into CF-based recommender models, acknowledging the significance of visual information in scenarios such as fashion-related recommendations. Furthermore, the utilization of visual features helps address persistent issues in traditional recommender systems, such as data sparsity and cold starts, leading to performance improvements.

In contrast to VBPR, which directly uses pre-extracted CNN visual features, DVBPR~\cite{kang2017visually} takes a different approach to handling image information. Specifically, Kang et al.~\cite{kang2017visually} adopt an end-to-end framework for DVBPR to train a CNN model with raw image input for visual feature extraction and simultaneously train the recommender model. ImRec~\cite{neve2020imrec} suggests leveraging reciprocal information between user groups through the use of image features. Chen et al.~\cite{chen2017attentive} propose ACF, which incorporates the attention mechanism into the CF model. It comprises item-level and component-level attention. The item-level attention is strategically employed to identify the most representative items that characterize individual users, while the component-level attention aims to extract the most informative features from multimedia auxiliary information for each user.

\subsection{Adversarial Attack on Visually-Aware Recommender Systems}
While visually-aware recommender systems indeed improve recommendation performance and alleviate the cold start issue, they also introduce new threats to these systems. Tang et al.~\cite{tang2019adversarial} reveal a security threat caused by malicious alterations to item images and further propose a robustness-focused visually-aware recommender model, AMR. Yin et al.~\cite{yin2023securing} proposed a framework capable of maintaining recommendation performance by denoising adversarial perturbations from attacked images (e.g., FGSM~\cite{goodfellow2014explaining}, PGD~\cite{kurakin2018adversarial}), as well as detecting adversarial attacks. However, this work is constrained to defending against untargeted attacks under the assumption of a white-box attack. Merra et al.~\cite{merra2023denoise} introduced a novel method called AiD, which has the ability to remove adversarial perturbations from attacked images prior to their use in visually-aware recommender systems. To effectively remove perturbations from adversarial images, the AiD model requires a dataset consisting of clean images and their corresponding noised images for training. Alternatively, the defender must possess knowledge of the specific attack method used to generate the adversarial images in order to generate the necessary noised images for training the AiD model.

To date, adversarial approaches for the visual data in the domain of visually-aware recommender systems can be categorized into two main types: classifier-targeted attacks and ranker-targeted attacks.

In a classifier-targeted attack, the objective is to modify the predictions of item categories without directly impacting the ranking of items. TAaMR~\cite{di2020taamr} is noteworthy as a representative work in this category. Specifically, classifier-targeted attacks, such as TAaMR, generate adversarial images with the ability to deceive the image classifier, transitioning from the source class (e.g., bottles) to the target class (e.g., shoes), while maintaining visual consistency with the original image. However, there is an evident and inherent limitation to such classifier-targeted attacks, arising from the obligatory use of class labels. Technically, adversarial images generated by classifier-targeted attacks could be ineffective when the class categories for the source and target are the same. For instance, attacking a shoe image could be highly challenging if the target class is also a shoe.

In a ranker-targeted attack, adversarial images are purposefully crafted to directly perturb the ranker of recommender systems, aiming to either promote or demote items. To the best of our knowledge, AIP~\cite{liu2021adversarial} stands as the first and only work dedicated to ranker-targeted attacks within the context of visually-aware recommender systems. Specifically, AIP generates adversarial images by introducing an optimized perturbation to the target item's image, aiming to reduce the distance between the visual vectors of the target item and popular items. However, the effectiveness of the AIP attack is constrained by a loose perturbation scale and the selection of popular images. Additionally, noticeable image distortion occurs, making the adversarial image easily recognizable by users.

Therefore, both the existing attacks on the visual data of the visually-aware recommender systems are unable to expose the genuine vulnerability of the system.

\subsection{Diffusion Model}
Diffusion models are well-known for their ability to generate high-quality data, particularly in the computer vision domain. Drawing inspiration from non-equilibrium thermodynamics~\cite{sohl2015deep}, the diffusion model operates with a different mechanism compared to other commonly used generative models such as GANs~\cite{goodfellow2020generative} and VAEs~\cite{kingma2013auto}. Specifically, the diffusion model is composed of two Markov chains that represent the forward and reverse processes, respectively. In the forward process, noises randomly sampled from the Gaussian distribution are gradually added to the original data. In the reverse process, the model predicts and removes the noises to generate new data ~\cite{yang2023diffusion}.

DDPM~\cite{ho2020denoising} represents a pioneering effort in adopting the diffusion model for generating high-quality images. However, due to the unpredictability of noise denoising in the reverse process, the content of the generated image becomes random. To address this, Dhariwal et al.~\cite{dhariwal2021diffusion} propose a guided diffusion model that introduces a condition to constrain noise removal, thereby controlling the data distribution of the generated image. In addition to applications in the computer vision domain, ~\cite{li2023diffurec, du2023sequential, wang2023diffusion} are works that utilize the diffusion model in the context of recommendations. Yuan et al.~\cite{yuan2023manipulating1} present the first work employing the diffusion model for enhancing the security of federated recommender systems.

\section{Preliminaries}\label{preliminaries}
In this section, we provide a discussion of the preliminaries of our research by introducing the base visually-aware recommender systems and the adversarial goals and prior knowedge employed by IPDGI.

\subsection{Base Visually-Aware Recommender Systems}
In broad terms, visually-aware recommender systems are those that incorporate visual features into preference predictions. We have selected three representative visually-aware recommender systems as base models to evaluate the effectiveness and imperceptibility of the IPDGI attack. The chosen models include VBPR~\cite{he2016vbpr}, DVBPR~\cite{kang2017visually}, and AMR~\cite{tang2019adversarial}, each distinguished by a unique mechanism.

\subsubsection{VBPR}
Visual Bayesian Personalized Ranking (VBPR) is a pioneering visually-aware recommender system that leverages images as auxiliary information for predicting users' preferences, specifically designed to alleviate the cold start issue. It is a Bayesian Personalized Ranking (BPR)-based method~\cite{rendle2012bpr} extended to incorporate visual features into the latent features of users and items. More precisely, VBPR integrates the Convolutional Neural Network (CNN)~\cite{he2016deep, simonyan2014very} pre-extracted visual features of items with the latent (non-visual) features to form the item representation. The predictive model of VBPR can be articulated as follows:
\begin{equation} \label{eq_vbpr}
    y_{u,i} = \chi + \eta_u + \eta_i + \lambda_u^T\lambda_i + \phi_u^T\phi_i + \eta_v^T{f_i}
\end{equation}
\begin{equation} \label{eq_vbpr_visual_trans}
    \phi_i = \mathbf{E}f_i
\end{equation}
where $y_{u,i}$ is the predicted score that user $u$ given to item $i$; $\chi$ is the global offset; $\eta_u$ and $\eta_i$ are the biases associated with user $u$ and item $i$, respectively; $\lambda_u$ and $\lambda_i$ are the latent features for user $u$ and item $i$, respectively; and $\phi_u$ and $\phi_i$ represent the visual features for user $u$ and item $i$, respectively. The visual feature for item $i$ is denoted as $f_i$, which is obtained through CNN extraction. Furthermore, the visual bias is represented by $\eta_v^T{f_i}$. Due to the high dimensionality of the extracted image feature $f_i$, it cannot be directly used as the visual feature $\phi_i$. Alternatively, He et al.~\cite{he2016vbpr} proposed a learnable embedding $\mathbf{E}$ to transform the extracted feature $f_i$ from the CNN space into a lower-dimensional visual space, as described by Eq.~\ref{eq_vbpr_visual_trans}. In addition to its efficacy in mitigating the cold start issue, VBPR has demonstrated enhanced recommendation performance, along with notable transparency and interpretability in the recommendation process.

\subsubsection{DVBPR}
Deep Visual Bayesian Personalized Ranking (DVBPR) constitutes a recommender system built upon the foundation of VBPR, specially tailored for fashion recommendation scenarios. DVBPR distinguishes itself from the original VBPR systems through its unique approach to leveraging item images. According to~\cite{lei2016comparative, veit2015learning}, DVBPR~\cite{kang2017visually} employs a CNN model to directly extract visual features from item images rather than relying on pre-extracted CNN visual features. Specifically, DVBPR utilizes an end-to-end framework that concurrently employs a CNN model to extract visual features and a recommender model to learn user latent factors. Kang et al.~\cite{kang2017visually} argue that discarding item bias and non-visual latent factors is justified, as the remaining terms adequately capture implicit factors under the end-to-end approach of extracting visual features. Consequently, the preference predictor of DVBPR can be expressed as:
\begin{equation} \label{eq_dvbpr_cnn}
    y_{u,i} = \chi + \eta_u + \phi_u^T \Psi(\mathbf{X}_i)
\end{equation}
where $\Psi(\mathbf{X}_i)$ represents the CNN model $\Psi(\cdot)$ with the item $i$'s image $\mathbf{X}_i$. Similar to VBPR, the recommender model of DVBPR is also BPR-based, with the primary objective of optimizing rankings through the consideration of triplets $(u,i,j)\in\mathcal{D}$. This can be defined as:
\begin{equation} \label{eq_bpr}
\begin{aligned}
    &y_{u,i,j} = y_{u,i} - y_{u,j},\\
    &\mathrm{where}\; \mathcal{D} = \{(u,i,j)\vert u\in\mathcal{U} \wedge i\in\mathcal{I}_u^+ \wedge j\in\mathcal{I}\backslash\mathcal{I}_u^+\}
\end{aligned}
\end{equation}
In Eq.~\ref{eq_bpr}, $\mathcal{U}$ and $\mathcal{I}$ represent the sets of users and items, respectively. For an item $i\in\mathcal{I}_u^+$, it signifies an item that the user $u$ has interacted with or expressed interest in, while $j\in\mathcal{I}\backslash\mathcal{I}_u^+$ represents an item that the user $u$ has not interacted with or expressed interest in. Moreover, following the BPR expression, the global bias $\chi$ and user bias $\eta_u$ can be eliminated due to the cancellation between $y_{u,i}$ and $y_{u,j}$. Consequently, the DVBPR predictor (see Eq.~\ref{eq_dvbpr_cnn}) can be further simplified, yielding the final form of the preference predictor as:
\begin{equation} \label{eq_dvbpr_simplified}
    y_{u,i} = \phi_u^T \Psi(\mathbf{X}_i)
\end{equation}

\subsubsection{AMR}
Adversarial Multimedia Recommendation (AMR) is a visually-aware recommender system with a focus on robustness. It is built upon VBPR and utilizes the same preference predictor (see Eq.~\ref{eq_vbpr}). Specifically, AMR integrates the VBPR recommender model with an adversarial training procedure to enhance model robustness. In this approach, adversarial perturbations are proactively introduced to the visual features of items during the recommender model training, as defined by:
\begin{equation}
    y_{u,i} = \chi + \eta_u + \eta_i + \lambda_u^T\lambda_i + \phi_u^T\cdot\mathbf{E}(f_i + \Delta_i) + \eta_v^T\cdot{(f_i + \Delta_i)}
\end{equation}
where $\Delta_i$ represents the adversarial perturbations optimized to exert the most significant influence, corresponding to the worst-case scenario, on the recommender model. The optimization process for these adversarial perturbations is detailed in Eq.~\ref{eq_amr_perturbations}.
\begin{equation} \label{eq_amr_perturbations}
\begin{aligned}
    &\Delta_i = L_{BPR}^{adv} = \underset{\Delta}{\arg\min} \sum_{(u,i,j)\in\mathcal{D}} -\ln\varsigma(y_{u,i}^{adv} - y_{u,j}^{adv}),\\
    &\mathrm{where}\; \lVert\Delta_i\rVert \leq \upsilon,\; i=1,...,\vert\mathcal{I}\vert;\; \lVert\Delta_j\rVert \leq \upsilon,\; j=1,...,\vert\mathcal{I}\vert
\end{aligned}
\end{equation}
Here, $\varsigma(\cdot)$ denotes the sigmoid function, $\lVert\cdot\rVert$ represents the $L_2$ norm, and $\upsilon$ is the magnitude to restrict the perturbations. As the AMR method involves a minimax game, perturbations $\Delta$ are learned to maximize the loss function of the recommender model, while simultaneously, the model parameters $\Theta$ are learned to minimize both the loss function and the adversary's loss (see Eq.\ref{eq_amr_optimization}).
\begin{equation} \label{eq_amr_optimization}
\begin{aligned}
    \Theta &= \underset{\Theta}{\arg\min}\; L_{BPR} + \varphi L_{BPR}^{adv}\\
    &= \underset{\Theta}{\arg\min} \sum_{(u,i,j)\in\mathcal{D}} -\ln\varsigma(y_{u,i} - y_{u,j}) - \varphi\ln\varsigma(y_{u,i}^{adv} - y_{u,j}^{adv}) + \tau\lVert\Theta\rVert^2
\end{aligned}
\end{equation}
where $\tau$ regulates the strength of $L_2$ regularization on model parameters, and $\varphi$ is a hyper-parameter controlling the impact of the adversary on model optimization. Specifically, the adversary has no impact if $\varphi$ is set to 0. This dual learning approach enhances the model's robustness to adversarial perturbations in multimedia content, resulting in a diminished impact on the model's predictions.

\subsection{Adversarial Approaches for Visually-aware Recommender Systems}
\textbf{Adversarial Goal.}
The primary objective of employing adversarial images in this paper is to promote target items within the top-K ranker of visually-aware recommender systems, i.e., enhance the exposure rate of the target items. Additionally, the adversarial images should closely resemble the visual appearance of the original images, ensuring they appear natural to users to maintain stealthiness while preserving the effectiveness of the attack. Furthermore, the overall recommendation performance should not be significantly compromised when the recommender system is under attack, even in the worst-case scenario.

\textbf{Adversarial Prior Knowledge.}
In this paper, we explore the vulnerabilities of visually-aware recommender systems within a real-world and practical context. Therefore, we assume that adversaries have minimal internal knowledge of the system. The only prior knowledge we attribute to adversaries is familiarity with the visual feature extraction model employed in the target visually-aware recommender system, denoted as $\Psi$.

\section{Our Approach}\label{our_approach}
\begin{figure*}[ht]
    \centering
    \includegraphics[scale=0.1]{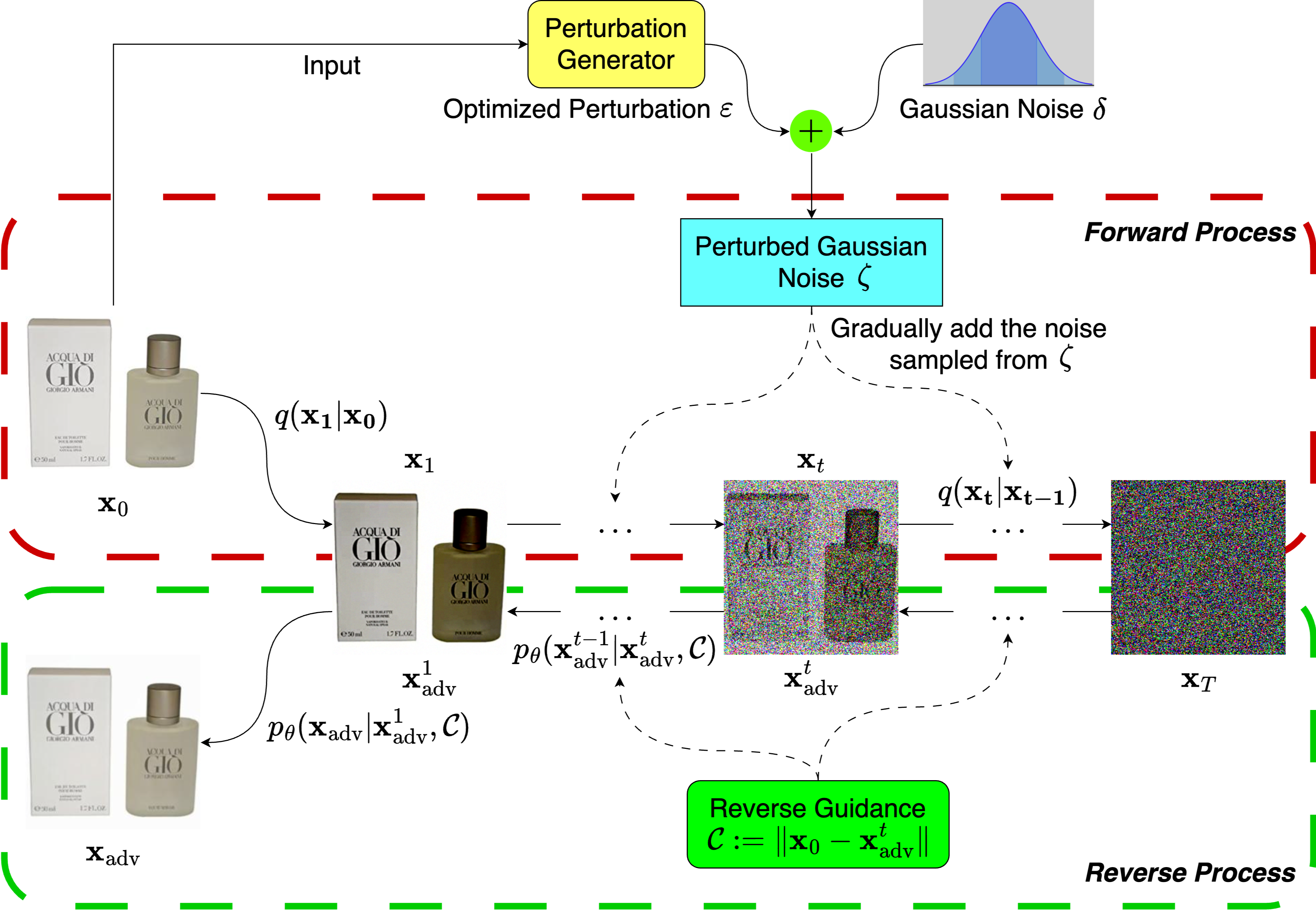}
    \caption{IPDGI Overview}
    \label{fig_ipdgi}
\end{figure*}

In this section, we delve into the details of our approach, the Item Promotion by Diffusion Generated Images (IPDGI) attack. Figures~\ref{fig_ipdgi} and ~\ref{fig_perturbation_gen} illustrate the overview and perturbation generation process of IPDGI, respectively. Additionally, Algorithm ~\ref{algo_ipdgi} is the pseudocode of IPDGI attack.

\subsection{Base Diffusion Model}
Diffusion models such as DDPM~\cite{ho2020denoising} are probabilistic generative models consisting of two processes: the forward process and the reverse process, both of which can be represented as Markov chains. Initially, in the forward process, noise is randomly sampled from a Gaussian distribution and added to the input image for $T$ steps, gradually transforming the original image into complete Gaussian noise. Subsequently, in the reverse process, the diffusion model is trained to iteratively reverse the noise image generated from the forward process, recovering to a clean image that shares the same data distribution as the input image.

\subsubsection{Forward Process}\label{forward_process}
The forward process, a.k.a. the diffusion process, is a Markov chain with the goal of transforming the data distribution of the input image into a Gaussian distribution by iteratively adding the Gaussian noise to it. Formally, according to the chain rule of probability and the property of Markov chains, the forward process generates the noisy samples $\mathbf{x}_1, \mathbf{x}_2,..., \mathbf{x}_T$, as follows: 
\begin{equation} \label{eq_1}
    q(\mathbf{x}_1, ..., \mathbf{x}_T \vert \mathbf{x}_0) = \prod_{t=1}^{T} q(\mathbf{x}_t \vert \mathbf{x}_{t-1})
\end{equation}

Here, $T$ denotes the number of diffusion steps, and the transformation process $q(\mathbf{x}_t \vert \mathbf{x}_{t-1})$ turns the data distribution of $q(\mathbf{x}_0)$ into a tractable prior distribution by gradually adding Gaussian noise. In DDPM, the $q(\mathbf{x}_t \vert \mathbf{x}_{t-1})$ process has the following representation:
\begin{equation} \label{eq_2}
    q(\mathbf{x}_t \vert \mathbf{x}_{t-1}) = \mathcal{N}(\mathbf{x}_t; \sqrt{1 - \beta_{t}} \mathbf{x}_{t-1}, \beta_{t} \mathbf{I}), \; \forall t \in \{1, ..., T\} \;,
\end{equation}
where $\mathcal{N}(\mathbf{x}_t; \sqrt{1 - \beta_{t}} \mathbf{x}_{t-1}, \beta_{t}\mathbf{I})$ can be expressed in the general form, $\mathcal{N}(\mathbf{x}_t; \mu, \sigma^2)$, indicating that $\mathbf{x}_t$ is generated by the Gaussian distribution with mean $\mu$ and variance $\sigma$. $\beta_t$ represents the noise at step $t$, which is pre-scheduled. Commonly, the scheduled noise $\beta$ can be generated in the manner of either linear, cosine, or square-root. According to Ho et al. ~\cite{ho2020denoising}, we can simplify Eq.~\ref{eq_2} by directly calculating $\mathbf{x}_t$ conditioned on $\mathbf{x}_0$ at an arbitrary diffusion step with the following transformation:
\begin{equation} \label{eq_3}
\begin{aligned}
        &q(\mathbf{x}_t \vert \mathbf{x}_0) = \mathcal{N}(\mathbf{x}_t; \sqrt{\bar{\alpha_t}} \mathbf{x}_0, (1 - \bar{\alpha_t}) \mathbf{I})\\
        &\alpha_t := 1 - \beta_t,\quad \bar{\alpha_t} := \prod_{s=1}^{t} \alpha_s
\end{aligned}
\end{equation}

Then, with re-parameter tricks, $\mathbf{x}_t$ can be computed as follows: 
\begin{equation} \label{eq_4}
    \mathbf{x}_t = \sqrt{\bar{\alpha_t}} \cdot \mathbf{x}_0 + \sqrt{1 - \bar{\alpha_t}} \cdot \delta, \; \mathrm{where} \; \delta \sim \mathcal{N}(0, \mathbf{I})
\end{equation}

\subsubsection{Reverse Process}
Unlike the forward process, which gradually corrupts the original image $\mathbf{x}_0$ into Gaussian noise with schedules, the reverse process is a trainable Markov chain that approximates $\mathbf{x}_{0}$ by predicting and removing noise from $\mathbf{x}_T$. Formally, the learnable reverse process from step $T$ to 0 can be defined as:
\begin{equation} \label{eq_5}
    p_\theta(\hat{\mathbf{x}}_0, ..., \hat{\mathbf{x}}_t, ..., \hat{\mathbf{x}}_{T-1} \vert \mathbf{x}_T) = \prod_{t=1}^{T} p_\theta(\hat{\mathbf{x}}_{t-1} \vert \hat{\mathbf{x}}_t) \;,
\end{equation}
where $\theta$ denotes the model parameters. The learnable reverse process $p_\theta(\hat{\mathbf{x}}_{t-1} \vert \hat{\mathbf{x}}_t)$ takes the diffused input $\hat{\mathbf{x}}_t$ and its corresponding time embedding $t$ to predict the mean $\mu_\theta(\hat{\mathbf{x}}_t,t)$ and variance $\Sigma_\theta(\hat{\mathbf{x}}_t,t)$, as shown in Eq.~\ref{eq_6}. 
\begin{equation} \label{eq_6}
    p_\theta(\hat{\mathbf{x}}_{t-1} \vert \hat{\mathbf{x}}_t) = \mathcal{N}(\hat{\mathbf{x}}_{t-1}; \mu_\theta(\hat{\mathbf{x}}_t,t), \Sigma_\theta(\hat{\mathbf{x}}_t,t))
\end{equation}
\begin{equation} \label{eq_7}
    \mu_\theta(\hat{\mathbf{x}}_t,t) = \frac{1}{\sqrt{\alpha_t}} (\hat{\mathbf{x}}_t - \frac{\beta_t}{\sqrt{1 - \bar{\alpha_t}}} z_\theta(\hat{\mathbf{x}}_t,t))
\end{equation}

In practice~\cite{ho2020denoising}, the variance $\Sigma_\theta(\hat{\mathbf{x}}_t,t)$ is treated as constant values to reduce the training complexity. As a result, the objective of the reverse process is simplified to reduce the distance between the real noise $\mathbf{z}_t$ and the predicted noise $z_\theta(\hat{\mathbf{x}}_t,t)$ at the random step $t$, as defined in Eq.~\ref{eq_8}: 
\begin{equation} \label{eq_8}
    \mathcal{L}_{simple} = \mathbb{E}_{t \sim [1,T]} \mathbb{E}_{\mathbf{x}_0 \sim p(\mathbf{x}_0)} \mathbb{E}_{\mathbf{z}_t \sim \mathcal{N}(0,\mathbf{I})} \lVert{\mathbf{z}_t - z_\theta(\hat{\mathbf{x}}_t,t)}\rVert^2
\end{equation}
where $\mathbf{x}_0 \sim p(\mathbf{x}_0)$ is normal image sampled from training data.

\subsection{Guided Diffusion for Adversarial Item Promotion}\label{guided_diffusion}
While Diffusion models~\cite{ho2020denoising} exhibit the ability to produce high-quality synthetic images, they inherently introduce diversity in the generated outputs. In other words, the images generated by a diffusion model can be random and divergent from the original input image. In the context of our visual attack on recommender systems, this randomness should be avoided as our adversarial goal is to promote items while maintaining high similarity between the content of adversarial images and their corresponding originals. Essentially, this randomness issue stems from the fact that sample generation in the reverse process occurs without conditional constraints. Inspired by guided diffusion ~\cite{dhariwal2021diffusion} in the computer vision domain that used a classifier as guidance for the reverse process, we have adopted a conditional constraint into the reverse process to guide diffusion sampling at each step $t$. Specifically, the reverse process in Eq.~\ref{eq_6} is transformed to Eq.~\ref{eq_9} with a condition:
\begin{equation} \label{eq_9}
    p_{\theta}(\mathbf{x}_{\mathrm{adv}}^{t-1} \vert \mathbf{x}_{\mathrm{adv}}^t,\mathcal{C}) = p_\theta(\mathbf{x}_{\mathrm{adv}}^{t-1} \vert \mathbf{x}_{\mathrm{adv}}^t) p(\mathcal{C}) \;,
\end{equation}
where $\mathcal{C}$ denotes the conditional constraint. Here, $p(\mathcal{C})$ is defined as $p(\mathbf{x}_0 \vert \mathbf{x}_{\mathrm{adv}}^t)$, since we aim to force the reverse steps considering more about the original input image. Intuitively, $p(\mathbf{x}_0 \vert \mathbf{x}_{\mathrm{adv}}^t)$ can be interpreted as ``The possibility of recovering to the original image $\mathbf{x}_{0}$ based on the current reversed image $\mathbf{x}_{\mathrm{adv}}^{t}$''.
Then, according to ~\cite{sohl2015deep, dhariwal2021diffusion}, we have the approximation of Eq.~\ref{eq_9} as:
\begin{equation} \label{eq_10}
\begin{aligned}
    \log p_{\theta}(\mathbf{x}_{\mathrm{adv}}^{t-1} \vert \mathbf{x}_{\mathrm{adv}}^t,\mathcal{C}) &\approx \log p_{\theta}(\mathbf{x}_{\mathrm{adv}}^{t-1} \vert \mathbf{x}_{\mathrm{adv}}^t)p(\mathcal{C}) \\
    &\approx \log p_{\theta}(\mathbf{x}_{\mathrm{adv}}^{t-1} \vert \mathbf{x}_{\mathrm{adv}}^t)p(\mathbf{x}_0 \vert \mathbf{x}_{\mathrm{adv}}^t) \\
    &\approx \log(z)
\end{aligned}
\end{equation}
\begin{equation} \label{eq_11}
    z \sim \mathcal{N}(\mu_\theta(\mathbf{x}_{\mathrm{adv}}^t,t) + \sigma_t^2 \nabla_{\mathbf{x}_{\mathrm{adv}}^t}\log p(\mathbf{x}_0 \vert \mathbf{x}_{\mathrm{adv}}^t), \sigma_t^2\mathbf{I}) \;,
\end{equation}

Further, we have opted for the Mean Squared Error (MSE) loss as the condition:
\begin{equation} \label{eq_12}
    p(\mathbf{x}_0 \vert \mathbf{x}_{\mathrm{adv}}^t) = \exp (\xi \lVert \mathbf{x}_0 - \mathbf{x}_{\mathrm{adv}}^t \rVert) \;,
\end{equation}
where $\xi$ denotes the guidance scale.
This design can effectively guide the reverse process to generate images that are similar to the original images, particularly in terms of pixel values. 

Finally, we define the reverse process of guided diffusion for adversarial sample as:
\begin{equation} \label{eq_13}
    p_{\theta}(\mathbf{x}_{\mathrm{adv}}^{t-1} \vert \mathbf{x}_{\mathrm{adv}}^t,\mathcal{C}) = \mathcal{N}(\mathbf{x}_{\mathrm{adv}}^{t-1}; \mu_\theta(\mathbf{x}_{\mathrm{adv}}^t,t) + \xi \cdot \sigma_t^2 \nabla_{\mathbf{x}_{\mathrm{adv}}^t} \lVert \mathbf{x}_0 - \mathbf{x}_{\mathrm{adv}}^t \rVert, \sigma_t^2\mathbf{I})
\end{equation}

\subsection{Perturbations for Adversarial Sample} \label{perturbations}
We utilize perturbations to generate adversarial images for the target item, typically the unpopular item. These adversarial images are intended to boost the ranking of the target item within the top-K recommendations. The fundamental idea of our attack is to add human-imperceptible noise to the target item image to shrink the distance of our target item's visual feature vector with the popular items' visual feature vectors: 
\begin{equation} \label{eq_14}
    \underset{\varepsilon}{\arg\min} \; \lVert \Psi(\mathbf{x}_{\mathrm{ref}}) - \Psi(\mathbf{x}_{\mathrm{adv}}) \rVert_2, \; \mathrm{where} \; \mathbf{x}_{\mathrm{adv}} = \mathbf{x}_i + \varepsilon, \; \varepsilon \sim \mathcal{N}(0,\mathbf{I}) \;,
\end{equation}
where $\mathbf{x}_{\mathrm{ref}}$, $\mathbf{x}_i$, and $\varepsilon$ represent the reference image, target image, and perturbation, respectively. $\Psi(\cdot)$ refers to the image feature extraction model~\cite{simonyan2014very,he2016deep} used in visually-aware recommender system.

In order to optimize Eq.~\ref{eq_14}, the first step is to find an appropriate popular item's image as the reference image. According to Liu et al.~\cite{liu2021adversarial}, the reference image is commonly chosen from the image of a popular item. This choice is made because the AIP attack operates by shifting the image space of the target item's image closer to that of the reference image through carefully designed perturbations. However, the selection of a reference image is crucial to ensure the attack's effectiveness and stealthiness since the objects in different images vary. Specifically, if the reference image has significant semantic differences from the target item's image, the resulting adversarial sample may become noticeably distorted as a larger $\varepsilon$ is needed to align $\Psi(\mathbf{x}_{\mathrm{adv}})$ with $\Psi(\mathbf{x}_{\mathrm{ref}})$. 

\begin{figure*}
    \centering
    \includegraphics[scale=0.1]{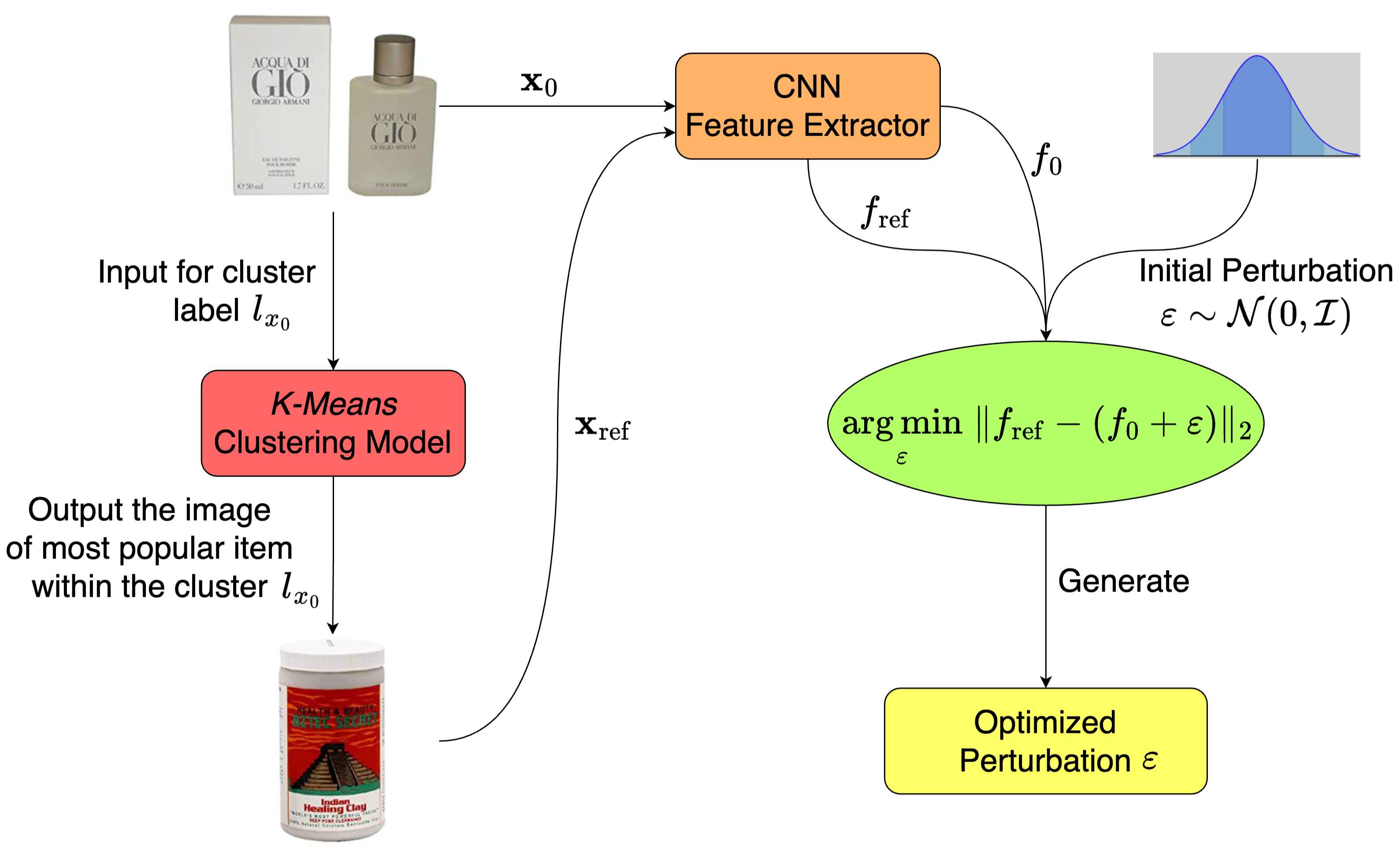}
    \caption{IPDGI Perturbation Generator}
    \label{fig_perturbation_gen}
\end{figure*}

In this paper, the reference image selection is as follows. We first employ \textit{k-means} cluster analysis on the images within the dataset, aiming to categorize images into different clusters based on their visual feature information. Then, we choose the image of the most popular items whose feature vectors are in the same cluster associated with our target items as the reference image, thereby minimizing the image semantic differences between them. Figure~\ref{fig_perturbation_gen} illustrates the perturbation generation process of IPDGI.

\subsection{Adversarial Samples Generation}
In Section~\ref{perturbations}, we discussed how to create perturbations to promote target items. Although we reduce the perturbation scales by selecting semantic-nearest reference images, customers may still discern adversarial images resulting from this visual attack, given that the perturbations are directly applied to regular images. To further improve the stealthiness of our attack, our IPDGI incorporates the perturbation within the diffusion model. This strategic choice is motivated by the proven efficacy of the diffusion model in producing high-quality synthetic images that closely resemble real-world counterparts, thus further refining the stealthiness of our approach. The details of combining attack perturbations and diffusion models are as follows. 

We first generate an adversarial perturbation $\varepsilon$ for the image of the target item following Eq.~\ref{eq_14}. We then combine it with the Gaussian noise $\delta$ from the base diffusion model (refer to Section~\ref{forward_process}). As illustrated in the perturbation generator process of IPDGI in Figure~\ref{fig_perturbation_gen}, the optimized adversarial perturbation is initially sampled from the Gaussian distribution, allowing it to naturally fuse with the Gaussian noise $\delta$, denoted as $\zeta := \varepsilon + \delta$ (see Figure~\ref{fig_ipdgi}). At this point, we have a perturbed Gaussian noise ready for the forward process of the guided diffusion model in the IPDGI. The forward process is further defined as follows:
\begin{equation} \label{eq_15}
    \mathbf{x}_t = \sqrt{\bar{\alpha_t}} \cdot \mathbf{x}_0 + \sqrt{1 - \bar{\alpha_t}} \cdot \zeta
\end{equation}

Since the pre-trained diffusion model we used being trained with normal images, i.e., the uncorrupted/undistorted images, the denoising within the reverse process tends to restore $\mathbf{x}_T$ (i.e., the perturbed Gaussian noise image generated through forward process) to the domain of clean images. Specifically, the reverse process can remove the noises that cause image distortion while preserving the perturbations in the image to deceive the top-K ranker of the recommender systems. The reverse process remains the same as the Eq.~\ref{eq_13} in section~\ref{guided_diffusion}.

\begin{algorithm}
\renewcommand{\algorithmicrequire}{\textbf{Input:}}
\renewcommand{\algorithmicensure}{\textbf{Output:}}
\caption{IPDGI: Item Promotion by Diffusion Generated Image} \label{algo_ipdgi}
    \begin{algorithmic}[1]
        \Require 
            \Statex $\mathbf{x}_0$: Original image of target item
            \Statex $\epsilon$: Perturbation magnitude
            \Statex $e$: Perturbation epochs
            \Statex $T$: Diffusion steps
            \Statex $\xi$: Guidance scale
            \Statex $\Psi$: Visual feature extractor
            \Statex $\kappa$: \textit{k-means} clustering model
        \Ensure 
            \Statex $\mathbf{x}_\mathrm{adv}$: Adversarial image of target item
        \Function{GeneratePerturbation}{$\mathbf{x}_0$, $\epsilon$, $e$, $\Psi$, $\kappa$}
            \State $\varepsilon \leftarrow$ Sample from $\mathcal{N}(0,\mathbf{I})$ \Comment{Initialise a perturbation by randomly sampling from a Gaussian distribution}
            \State $l_{\mathbf{x}_0} \leftarrow \kappa(\mathbf{x}_0)$ \Comment{Get the cluster label of the original image from $\kappa$}
            \State $\mathbf{x}_{\mathrm{ref}} \leftarrow$ Image of the most popular item within the cluster $l_{\mathbf{x}_0}$
            \For{$i \leftarrow 1$ to $e$}
                \State $f_{\mathrm{ref}} \leftarrow \Psi(\mathbf{x}_{\mathrm{ref}})$ \Comment{Visual feature of the reference image}
                \State $f_{\mathrm{x_i}} \leftarrow \Psi(\mathbf{x}_{i-1} + \varepsilon)$ \Comment{Visual feature of the target item's image containing the perturbation $\varepsilon$ at $i$ epoch}
                \State $\varepsilon \leftarrow \underset{\varepsilon}{\arg\min} \; \lVert f_{\mathrm{ref}} - f_{\mathrm{x_i}} \rVert_2$ \Comment{Refer to Eq.~\ref{eq_14}}
                \State $\varepsilon \leftarrow$ Resize the size of $\varepsilon$ according to perturbation magnitude/epsilon $\epsilon$
            \EndFor
            \State \Return $\varepsilon$
        \EndFunction
        \State $\varepsilon \leftarrow$ \Call{GeneratePerturbation}{$\mathbf{x}_0$, $\epsilon$, $e$, $\Psi$, $\kappa$}
        \State $\delta \leftarrow$ Sample from $\mathcal{N}(0,\mathbf{I})$
        \State $\zeta \leftarrow \varepsilon + \delta$ \Comment{Perturbed Gaussian noise for the diffusion process}
        \State $\mathbf{x}_T \leftarrow$ Generate the complete Gaussian noise for $\mathbf{x}_0$ through the diffusion process \Comment{Refer to Eq.~\ref{eq_15}}
        \For{$i \leftarrow T$ to $1$}
            \State $\mathbf{\bm{\mu}}$, $\Sigma \leftarrow \mu_\theta(\mathbf{x}_{\mathrm{adv}}^t,t)$, $\sigma_t^2$
            \State $\mathbf{x}_{\mathrm{adv}}^{t-1} \leftarrow$ Sample from $\mathcal{N}(\bm{\mu} + \xi \cdot \Sigma \nabla_{\mathbf{x}_{\mathrm{adv}}^t} \lVert \mathbf{x}_0 - \mathbf{x}_{\mathrm{adv}}^t \rVert, \Sigma\mathbf{I})$ \Comment{Refer to Eq.~\ref{eq_13}}
        \EndFor
        \State $\mathbf{x}_\mathrm{adv} \leftarrow \mathbf{x}_{\mathrm{adv}}^{t-1}$
        \State \Return $\mathbf{x}_\mathrm{adv}$
    \end{algorithmic}
\end{algorithm}

\section{Experiments}\label{experiments}
In this section, we conduct extensive experiments to answer the following research questions (RQs):
\begin{itemize}
    \item \textbf{RQ1:} How is the attack effectiveness of our proposed IPDGI?
    \item \textbf{RQ2:} How is the attack stealthiness of our proposed IPDGI?
     \item \textbf{RQ3:} How is the attack impact of IPDGI on normal recommendation performance?
    \item \textbf{RQ4:} How do the hyper-parameters influence the effectiveness and imperceptibility of the IPDGI?
\end{itemize}

\subsection{Experimental Settings}
\subsubsection{Dataset} \label{dataset}
We conduct experiments on two real-world recommendation datasets, namely \textit{Amazon Beauty} and \textit{Amazon Baby}, both derived from the Amazon website ~\cite{mcauley2015image}. We chose these two datasets because the visual signal is vital in influencing customers' final decisions in the two domains. In addition, these two datasets have proper sizes so that our experimental results could be easy to be reproduced by other researchers. For both datasets, we filter out those users and items with less than ten interactions (i.e., 10-core filtering) following ~\cite{qu2023semi,zhao2022revisiting}. After filtering, the \textit{Amazon Beauty} dataset consists of $8,787$ users, $1,480$ items, and $62,631$ user-item interactions. On the other hand, the \textit{Amazon Baby} dataset includes $6,158$ users and $1,009$ items, and $44,335$ user-item interactions. Each item has one corresponding image. Then, following the common settings in implicit feedback recommender models~\cite{yuan2023manipulating1,he2017neural,zhang2021graph}, we binarize the user-item ratings by transforming all the ratings contained in the dataset to $r_{ij}=1$ and negative instances are sampled with $1:4$ ratio~\cite{he2017neural,xia2022hypergraph} to train visually-aware recommender systems. Table~\ref{t_dataset} illustrates the statistics of these two datasets.

\begin{table}[ht]
    \centering
    \caption{Statistics of Amazon Beauty and Amazon Baby}
    \begin{tabular}{ccccc}
        \toprule
        \textbf{Dataset} &\textbf{User\#} &\textbf{Item\#} &\textbf{Interactions\#} &\textbf{Sparsity} \\
        \midrule
        Amazon Beauty &$8,787$ &$1,480$ &$62,631$ &$99.52\%$ \\
        Amazon Baby &$6,158$ &$1,009$ &$44,335$ &$99.29\%$ \\
        \bottomrule
    \end{tabular}
    \label{t_dataset}
\end{table}

\subsubsection{Evaluation Protocol}
We employ the standard leave-one-out protocol~\cite{magnusson2019bayesian,he2017neural,zhang2022pipattack} to establish the training and testing data for each user. Specifically, for each user, we leave the last interacted item as the test item, while the remaining interacted items are utilized for training. In addition, we also select the last interacted item in the training data for validation during each training epoch. In order to simulate a more real attack scenario, we choose a set of unpopular items as target items to promote in the top-K recommendation. 
Following the approaches outlined in ~\cite{he2016vbpr, kang2017visually, tang2019adversarial}, we initially train the chosen base visually-aware recommender systems with uncorrupted images. Subsequently, we substitute the images of target items with the generated adversarial images to improve their rankings in the top-K recommendations. To analyze the performance of our attack method, we evaluate it from three perspectives: the attack effectiveness, the attack imperceptibility, and the recommendation accuracy. We employ the Exposure Rate at Rank K (ER@K)~\cite{zhang2022pipattack,yuan2023manipulating} and Normalized Discounted Cumulative Gain at K (NDCG@K) to measure the attack effectiveness and the recommendation accuracy in the top-K recommendation, respectively.  Consequently, the impact of attacks can be measured by the metric value differences between before and after the integration of adversarial images. In other words, a greater improvement in ER@K signifies a more effective attack, whereas a less decrease in NDCG@K indicates a more subtle impact on the recommendation accuracy caused by the attack. When calculating ER@K and NDCG@K, we rank all items. Therefore, our paper's ER@K and NDCG@K values look much lower than those calculated based on a small portion of randomly selected items (e.g., 100 randomly sampled negative items)~\cite{he2016vbpr,kang2017visually,tang2019adversarial,liu2021adversarial}. In addition, we adopt the Fréchet Inception Distance (FID)~\cite{heusel2017gans} metric to evaluate the quality of adversarial images. A smaller FID score suggests a higher similarity between the adversarial image and the original one, indicating a more imperceptible attack. To quantify the improvement or difference between two values, such as ``No Attack'' and IPDGI, we employ Eq.~\ref{eq_improvement_difference} for the calculation.
\begin{equation}\label{eq_improvement_difference}
    Improvement = \frac{\omega^{'} - \omega}{\omega} \times 100
\end{equation}
Here, $\omega$ and $\omega^{'}$ denote the original and new/updated ER@K values, respectively. For example, in the ``a vs. b'' scenario presented in Table~\ref{t_hr}, we have $\omega \leftarrow \mathrm{a}$ and $\omega^{'} \leftarrow \mathrm{b}$, indicating the changes between values ``a'' and ``b''.

\subsubsection{Baselines}
Since our attack method is ranker-targeted, we chose the same type of attack method as our baseline, the Adversarial Item Promotion (AIP) attack ~\cite{liu2021adversarial}. As far as our knowledge goes, the AIP attack is the sole ranker-targeted attack designed for visually-aware recommender systems. Similar to IPDGI, AIP adds noise to the target item's image to shrink the difference to popular items. However, it randomly chooses the popular item to optimize Eq.~\ref{eq_14}, resulting in unstable attack performance and obvious image distortion. In addition, we include ``No Attack'' to show the original ranking of the target items.

\subsection{Implementation Details}
In this section, we provide the implementation details of our experiments. The experiment pipeline is as follows. Firstly, we train the base models of visually-aware recommender systems, including VBPR~\cite{he2016vbpr}, DVBPR~\cite{kang2017visually}, and AMR~\cite{tang2019adversarial}, using uncorrupted item images. Subsequently, we calculate the average ER@K score of the target items and NDCG@K score of the testing data in the three base visually-aware recommender systems. We choose the unpopular items (with less than 20 interactions) as the target items to promote.  We apply adversarial attacks, including baseline attacks and our IPDGI, to generate corrupted images. These generated images are then utilized to substitute the regular images of unpopular items for promotional purposes within visually-aware recommender systems. The efficacy of these attacks is evaluated by comparing the variations in ER@K and NDCG@K scores and examining the FID values.

\subsubsection{Implementation of Visually-Aware Recommender Systems}
All visually-aware recommender models are implemented using PyTorch ~\cite{paszke2019pytorch}. The associated images of each item in the dataset used to train the base recommender system are uncorrupted. The size of user and item embeddings for all three models is set to 100~\cite{liu2021adversarial}. We use the pre-trained ResNet152 model ~\cite{he2016deep} as the image feature extractor $\Psi$ for all recommender models and the visual feature size is 2048. Following ~\cite{kang2017visually, tang2019adversarial, liu2021adversarial}, we train VBPR, DVBPR, and AMR for 2000, 50, and 2000 epochs, respectively. For the training optimizer, we adopt Adam ~\cite{kingma2014adam} with a learning rate set to 0.0001 and weight decay with a value of 0.001. After completing the training of these base visually-aware recommender systems, we choose the model with the lowest validation loss as the final model for each recommender system.

\subsubsection{Implementation of Attacks}\label{sec_attack_implement}
For the baseline attack AIP, we implement and execute it using the same settings in its original paper~\cite{liu2021adversarial}. Specifically, the perturbation training epochs are set to 5000 for each target image using Adam with a 0.001 learning rate as the optimizer. The maximum size of perturbation $\epsilon$ is set to 32. 

Our proposed IPDGI attack is a novel ranker-targeted attack based on the diffusion model, designed to promote a target item within the top-K ranker of a visually-aware recommender system in a stealthy manner. In IPDGI, we employ a $256\times256$ unconditional diffusion model weight~\footnote{\url{https://openaipublic.blob.core.windows.net/diffusion/jul-2021/256x256_diffusion_uncond.pt}}, which has been per-trained by ~\cite{dhariwal2021diffusion} on the ImageNet ~\cite{russakovsky2015imagenet} dataset.

\begin{table}[]
    \centering
    \small
    \setlength\tabcolsep{3.5pt}
    \caption{ER@$K$: Exposure Rate Comparison of AIP and IPDGI Attacks (where $K \in \lbrace5,10,20\rbrace$). The best-performing result on each visually-aware recommender system concerning each dataset is presented in boldfaced, while the second best is underscored.}
    \begin{tabular}{ccccccccc}
        \toprule
        \multirow{2}{*}{Dataset} &\multirow{2}{*}{\begin{tabular}[c]{@{}c@{}}Visually-Aware\\Recommender System\end{tabular}} &\multirow{2}{*}{ER@$K$} &(a) &(b) &(c) &\multicolumn{3}{c}{Improvement\;$\uparrow$} \\
        & & &No Attack &AIP &IPDGI &a vs. b &a vs. c &b vs. c \\
        \midrule
        \multirow{9}{*}{Amazon Beauty} 
            &\multirow{3}{*}{VBPR} 
                &5 &\underline{0.0095} &0.0091 &\textbf{0.0149} &-4.21\% &56.84\% &63.74\%\\
                & &10 &0.0270 &0.0270 &\textbf{0.0273} &0\% &1.11\% &1.11\%\\
                & &20 &\underline{0.0670} &0.0666 &\textbf{0.0674} &-0.60\% &0.60\% &1.20\%\\
                \cline{2-9}
            &\multirow{3}{*}{DVBPR} 
                &5 &0.0153 &0.0153 &\textbf{0.0171} &0\% &11.76\% &11.76\%\\
                & &10 &0.0313 &\underline{0.0314} &\textbf{0.0336} &0.32\% &7.35\% &7.01\%\\
                & &20 &0.0639 &\underline{0.0640} &\textbf{0.0667} &0.16\% &4.38\% &4.22\%\\
                \cline{2-9}
            &\multirow{3}{*}{AMR} 
                &5 &\underline{0.0183} &0.0173 &\textbf{0.0187} &-5.46\% &2.19\% &8.09\%\\
                & &10 &0.0312 &0.0312 &\textbf{0.0317} &0\% &1.60\% &1.60\%\\
                & &20 &\underline{0.0651} &0.0650 &\textbf{0.0657} &-0.15\% &0.92\% &1.08\%\\
        \midrule
        \multirow{9}{*}{Amazon Baby} 
            &\multirow{3}{*}{VBPR} 
                &5 &\underline{0.0180} &0.0160 &\textbf{0.0187} &-11.11\% &3.89\% &16.88\%\\
                & &10 &\underline{0.0327} &0.0320 &\textbf{0.0330} &-2.14\% &0.92\% &3.13\%\\
                & &20 &0.0677 &\underline{0.0680} &\textbf{0.0697} &0.44\% &2.95\% &2.50\%\\
                \cline{2-9}
            &\multirow{3}{*}{DVBPR} 
                &5 &\underline{0.0170} &0.0168 &\textbf{0.0180} &-1.18\% &5.88\% &7.14\%\\
                & &10 &0.0338 &\underline{0.0340} &\textbf{0.0348} &0.59\% &2.96\% &2.35\%\\
                & &20 &0.0661 &\underline{0.0663} &\textbf{0.0696} &0.30\% &5.30\% &4.98\%\\
                \cline{2-9}
            &\multirow{3}{*}{AMR} 
                &5 &\underline{0.0157} &0.0153 &\textbf{0.0165} &-2.55\% &5.10\% &7.84\%\\
                & &10 &\underline{0.0317} &0.0307 &\textbf{0.0333} &-3.15\% &5.05\% &8.47\%\\
                & &20 &0.0710 &\underline{0.0717} &\textbf{0.0723} &0.99\% &1.83\% &0.84\%\\
        \bottomrule
    \end{tabular}
    \label{t_hr}
\end{table}

\subsection{The Attack Effectiveness of IPDGI (RQ1)}
In this paper, we evaluate the efficacy of an attack based on exposure rate (ER@K, where $K \in \lbrace5,10,20\rbrace$). We conduct the experiments on two datasets, namely Amazon Beauty and Amazon Baby, with three visually-aware recommender systems (i.e., VBPR, DVBPR, AMR). Table~\ref{t_hr} presents the outcomes of a comparative analysis evaluating the effectiveness of various attacks, encompassing the baseline AIP attack (column label ``b''), our proposed method IPDGI attack (column label ``c''), and the original ranking performance labeled as ``No Attack'' (column label ``a''). Within the table, the highest score for each dataset and corresponding visually-aware recommender system is presented in bold, while the second-best score is underlined. Additionally, we provide the relative improvement results for the comparisons between ``No Attack'' and the AIP attack (a vs. b), ``No Attack'' and the IPDGI attack (a vs. c), and the AIP attack and IPDGI attack (b vs. c).

Firstly, in Table~\ref{t_hr}, we investigate the effectiveness of the AIP attack. The most significant improvement achieved by the AIP attack is 0.99\%, observed in the ER@20 for AMR on Amazon Baby when compared to the ``No Attack''. In other scenarios, the AIP attack demonstrates improvements of DVBPR on both datasets when evaluated using the ER@10 and ER@20 metrics. However, when assessed with the ER@5 metric, the AIP attack fails to promote target items in any scenario. Notably, there is an 11.11\% decline (i.e., -11.11\%) in the ER@5 for VBPR on Amazon Baby compared to the ``No Attack''. Moreover, under the scenarios of ER@10 for VBPR on Amazon Beauty, ER@5 for VBPR on Amazon Beauty, and ER@10 for AMR on Amazon Beauty, the AIP attack achieves the same exposure rate as the ``No Attack''. The failure of item promotion by the AIP attack can be attributed to the ineffectiveness of the perturbations added to the original images. AIP consistently selects the most popular item's image as the reference for all target items, without employing a technique such as a clustering model to analyze semantic differences between images. As a result, the perturbations lead to a huge difference in the data distribution of the generated images compared to the originals, causing image distortion and rendering the attack ineffective. These observations show that the AIP attack demonstrates limited improvement or is even ineffective under certain scenarios when considering item promotion against the top-K ranker of visually-aware recommender systems.

Secondly, we assess the performance of our method, the IPDGI attack. As shown in Table~\ref{t_hr}, the IPDGI attack consistently outperforms the baseline (AIP attack) and the ``No Attack'' (original performance), effectively promoting target items across all scenarios. In the ER@5 scenario for VBPR on Amazon Beauty, the IPDGI attack demonstrated its most significant improvements, achieving a 56.84\% improvement when compared with the ``No Attack'' (a vs. c) and a 63.74\% improvement when compared to the AIP attack (b vs. c).  Even for the robustness-focused visually-aware recommender system AMR, the IPDGI attack achieved ER@5 improvements of 2.19\% and 8.09\% compared to the ``No Attack'' and the AIP attack on the Amazon Beauty dataset.  Similarly, on the Amazon Baby dataset, the IPDGI attack exhibited improvements over the ``No Attack'' and the AIP attack, amounting to 5.10\% and 7.84\%, respectively.

Based on these observations, we posit that our method, the IPDGI attack, has demonstrated its effectiveness in targeting the base visually-aware recommender systems and exhibits a significant improvement over the baseline attack method.

\begin{table}[]
    \centering
    \caption{FID Scores: Adversarial Images by AIP and IPDGI Attacks. Lower FID scores indicate a higher degree of similarity between the generated images and the original ones.}
    \begin{tabular}{cccc}
        \toprule
            Dataset &Attack Method &FID\;$\downarrow$ &Improvement \\
        \midrule
        \multirow{2}{*}{Amazon Beauty} 
            &AIP &114.55 &\multirow{2}{*}{87.56\%} \\
            &IPDGI &\textbf{14.25} \\
        \midrule
        \multirow{2}{*}{Amazon Baby} 
            &AIP &265.48 &\multirow{2}{*}{91.51\%} \\
            &IPDGI &\textbf{22.54} \\
        \bottomrule
    \end{tabular}
    \label{t_fid}
\end{table}

\subsection{The Attack Imperceptibility of IPDGI (RQ2)}
\begin{figure*}
    \centering
    \includegraphics[scale=0.14]{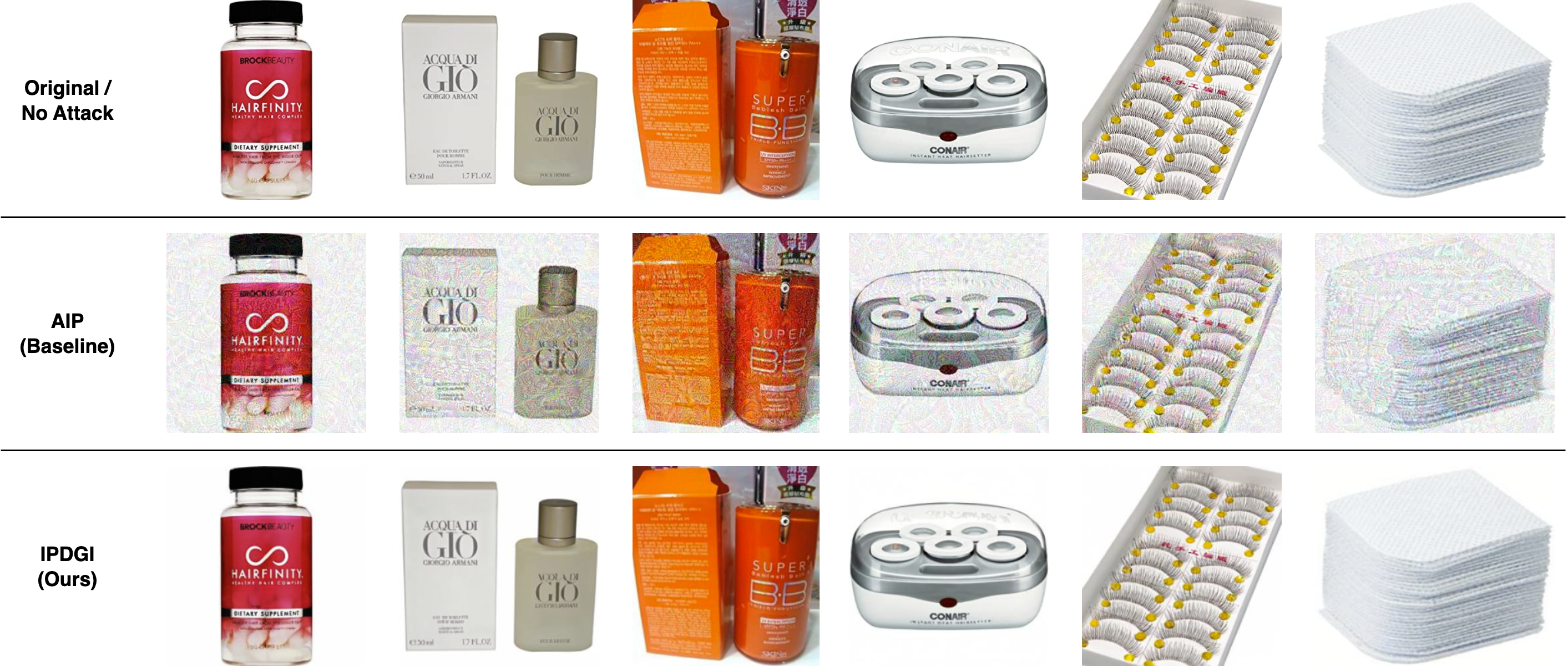}
    \caption{Comparisons of Attack Imperceptibility: Images Generated by the AIP Attack (Baseline) and the IPDGI Attack (Ours). The top row comprises uncorrupted images, the middle row displays images generated by the AIP attack, and the bottom row presents images generated by our IPDGI attack.}
    \label{fig_attack_imperceptibility_comparisons}
\end{figure*}

In this paper,  an attack's imperceptibility is measured by the consistency of adversarial images to the original ones. Notably, an attack capable of generating high-fidelity images becomes less noticeable to customers.  We utilize the FID metric to assess the similarity of adversarial images generated by the attack methods (i.e., AIP and IPDGI) with the original images, as shown in Table~\ref{t_fid}. A lower FID score signifies less difference between the real and generated images. For example, if FID is 0, there will be no difference between the two images. According to the results in Table~\ref{t_fid}, our attack method IPDGI outperforms the AIP attack on both datasets, achieving FID scores of 14.25 and 22.54 on Amazon Beauty and Amazon Baby, respectively. In contrast, the FID scores of the adversarial images generated by AIP are much higher, measuring 114.55 for Amazon Beauty and 265.48 for Amazon Baby, representing significantly lower image quality compared to IPDGI. Consequently, IPDGI demonstrates an 87.56\% and 91.51\% improvement over AIP in terms of image quality on Amazon Beauty and Amazon Baby, respectively. Figures~\ref{fig_image_comparisons} and ~\ref{fig_attack_imperceptibility_comparisons} present examples of adversarial images generated by different attacks. It is evident that the AIP images exhibit more severe distortions when compared to the original images, reflecting a loss of finer details. In contrast, the IPDGI images maintain high degrees of similarity with the originals.

\subsection{The Attack Impact on Normal Recommendation Performance (RQ3)}
In this experiment, we study the impact of the attack on the normal recommendation performance (i.e., recommendation accuracy).  For the side effects of attacks on the recommender system, we evaluate the NDCG@K value changes of the testing data before and after attacks. We utilize NDCG@K because it can directly reflect the item position change in a ranking list. Table~\ref{t_ndcg_perf} indicates the impact caused by the attacks on the visually-aware recommender systems. The values under ``No Attack'' indicate the average NDCG@K for all testing data when the recommender system is not corrupted, i.e., no adversarial images are used for the unpopular items. The columns ``AIP'' and ``IPDG'' show the average NDCG@K score after applying attacks. Specifically,  all target items (112 unpopular items on Amazon Beauty and 6 unpopular items on Amazon Baby) are corrupted by using adversarial images, representing the worst-case scenario. Overall, a smaller difference compared to the ``No Attack'' values (i.e., ``a vs. b'' and ``a vs. c'') implies that the attack has more subtle side effects. 

 As shown in Table~\ref{t_ndcg_perf}, we evaluate the average NDCG@K (where $K \in {5,10,20}$) for two different attack methods (i.e., AIP, IPDGI) across three visually-aware recommender systems on two datasets. For the Amazon Beauty dataset, the IPDGI attack results in slight declines in all of the base visually-aware recommender systems. This can be attributed to the successful promotion of target items, which are long-tail items. In other words, the adversarial images enable the target items to be exposed to a larger user audience.  In the offline evaluation setting, this would inevitably lead to a decline in the recommendation accuracy.  It should be noted here that the recommendation performance may not be compromised in the real-world online evaluation setting. 
  On the Amazon Baby dataset, both the AIP and IPDGI indicate no changes when compared to the ``No Attack'' scenario. This observation can be attributed to the small number of target items (only 6) in this dataset. Relative to the total number of items, this constitutes a relatively minor proportion. Consequently, the overall performances of the recommender systems are not affected by this limited amount of corrupted data or the attack.

By synthesizing the results from Tables~\ref{t_fid} and ~\ref{t_ndcg_perf}, we can infer that the adversarial images generated by the IPDGI attack are imperceptible in terms of image quality and have a minimal impact on the visually-aware recommender systems, even under the worst-case scenario.

\begin{table}[]
    \centering
    \small
    \setlength\tabcolsep{3.5pt}
    \caption{The side effects of Attacks on visually-aware recommender systems. A smaller ``Difference'' indicates subtle side effects. Notably, Amazon Beauty has 112 polluted items, whereas Amazon Baby has 6 polluted items.}
    \begin{tabular}{cccccccc}
        \toprule
        \multirow{2}{*}{Dataset} &\multirow{2}{*}{\begin{tabular}[c]{@{}c@{}}Visually-Aware\\Recommender System\end{tabular}} &\multirow{2}{*}{NDCG@$K$} &(a) &(b) &(c) &\multicolumn{2}{c}{Difference} \\
        & & &No Attack &AIP &IPDGI &a vs. b &a vs. c\\
        \midrule
        \multirow{9}{*}{Amazon Beauty} 
            &\multirow{3}{*}{VBPR} 
                &5 &0.0306 &0.0302 &0.0297 &-1.31\% &-2.94\%\\
                & &10 &0.0468 &0.0464 &0.0461 &-0.85\% &-1.50\%\\
                & &20 &0.0715 &0.0712 &0.0711 &-0.42\% &-0.56\%\\
                \cline{2-8}
            &\multirow{3}{*}{DVBPR} 
                &5 &0.0312 &0.0307 &0.0303 &-1.60\% &-2.88\%\\
                & &10 &0.0472 &0.0465 &0.0467 &-1.48\% &-1.06\%\\
                & &20 &0.0721 &0.0713 &0.0716 &-1.11\% &-0.69\%\\
                \cline{2-8}
            &\multirow{3}{*}{AMR} 
                &5 &0.0311 &0.0305 &0.0299 &-1.93\% &-3.86\%\\
                & &10 &0.0473 &0.0468 &0.0464 &-1.06\% &-1.90\%\\
                & &20 &0.0721 &0.0717 &0.0714 &-0.55\% &-0.97\%\\
        \midrule
        \multirow{9}{*}{Amazon Baby} 
            &\multirow{3}{*}{VBPR} 
                &5 &0.0323 &0.0323 &0.0323 &0\% &0\%\\
                & &10 &0.0485 &0.0485 &0.0485 &0\% &0\%\\
                & &20 &0.0735 &0.0735 &0.0735 &0\% &0\%\\
                \cline{2-8}
            &\multirow{3}{*}{DVBPR} 
                &5 &0.0325 &0.0325 &0.0325 &0\% &0\%\\
                & &10 &0.0487 &0.0487 &0.0487 &0\% &0\%\\
                & &20 &0.0737 &0.0737 &0.0737 &0\% &0\%\\
                \cline{2-8}
            &\multirow{3}{*}{AMR} 
                &5 &0.0334 &0.0334 &0.0334 &0\% &0\%\\
                & &10 &0.0495 &0.0495 &0.0495 &0\% &0\%\\
                & &20 &0.0743 &0.0743 &0.0743 &0\% &0\%\\
        \bottomrule
    \end{tabular}
    \label{t_ndcg_perf}
\end{table}

\subsection{The Effects of Hyper-Parameters on IPDGI (RQ4)}
In this section, we delve into the impact of varying hyper-parameters of IPDGI on the generation of adversarial images. IPDGI has four most important hyper-parameters: maximum perturbation scale $\epsilon$, perturbation optimization epochs $e$, diffusion steps $T$, and guidance scale $\xi$. Specifically, $\epsilon$ is the perturbation magnitude used to determine the strength of the perturbation for generating adversarial images. The epochs $e$ represent the number of iterations used to generate the perturbation. The diffusion steps $T$ indicate the number used in the forward and reverse processes of the diffusion model. The guidance scale $\xi$ is the factor controlling the guidance during the reverse process of the diffusion model. To facilitate analysis, when investigating a single hyper-parameter, the remaining three hyper-parameters were held constant at their default values: 16 for epsilon, 30 for the number of epochs, and 15 for both the diffusion steps and guidance scale. For each hyper-parameter, we considered five possible values for testing. 

\begin{figure}
    \centering
    \includegraphics[scale=0.6]{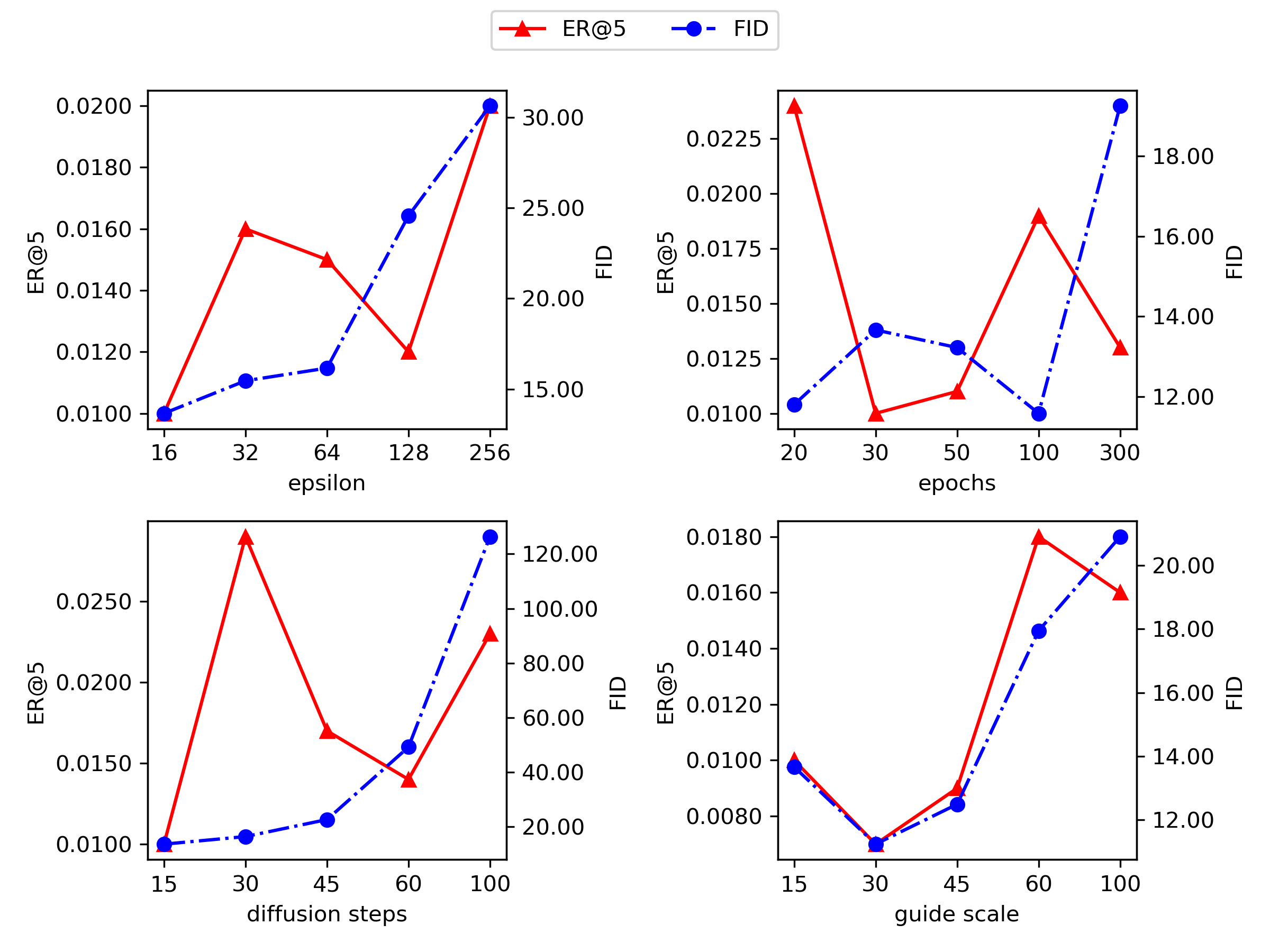}
    \caption{The analysis of Exposure Rate (ER@5) and image quality (FID) respect to different hyper-parameters (i.e., epsilon $\epsilon$, epochs $e$, purify steps $T$, and guide scale $\xi$)}
    \label{fig_hyperparams_fid}
\end{figure}

Figure~\ref{fig_hyperparams_fid} illustrates the impact on the effectiveness and imperceptibility of IPDGI resulting from variations in the hyper-parameter values. Each sub-figure depicts changes in the exposure rate (ER@5) and image quality (FID) corresponding to different hyper-parameter values, denoted by red triangles and blue dots, respectively. The left y-axis represents the exposure rate, while the right y-axis represents image quality. The x-axis of each sub-figure in Figure~\ref{fig_hyperparams_fid} indicates the testing values of the hyper-parameter.

\textbf{Impact of $\epsilon$.} 
As depicted in the top-left sub-figure of Figure~\ref{fig_hyperparams_fid}, an increment in the value of $\epsilon$ from 16 to 256 leads to an elevation in the FID score of the adversarial image, indicating a degradation in image quality. This outcome is expected, given that a larger $\epsilon$ corresponds to the introduction of stronger noise to the original images. However, it is noteworthy that the increase in $\epsilon$ does not necessarily result in an improvement in attack effectiveness. This suggests that a larger perturbation does not consistently yield better attack performance. For instance, at $\epsilon = 256$, the attack attains the highest ER@5 scores, while at $\epsilon = 32$ or $\epsilon = 64$, the ER@5 performance surpasses that at $\epsilon = 128$.

\textbf{Impact of $e$.} 
Based on the observations from the top-right sub-figure in Figure~\ref{fig_hyperparams_fid}, it is evident that the two highest ER@5 results are achieved at $e = 20$ and $e = 100$. Similarly, the two best FID scores are also obtained at these epochs (i.e., $e = 20$ and $e = 100$). From these findings, we contend that the judicious selection of the epoch is capable of generating well-optimized adversarial perturbations for the target image. Such perturbations enable the achievement of an improved exposure rate (i.e., a high ER@5 score) while simultaneously minimizing the impact on image quality (i.e., a low FID score).

\textbf{Impact of $T$.}
The diffusion step significantly influences the performance of IPDGI in both ER@5 and FID. Concerning the impact on image quality with different diffusion steps, as illustrated in the bottom-left sub-figure of Figure~\ref{fig_hyperparams_fid}, it is evident that larger diffusion steps result in lower image quality. This is attributed to the larger steps in the diffusion processes, increasing the likelihood of the generated image deviating from the original. In the analysis of attack effectiveness, we observed that for diffusion steps $T = 30$ and $T = 100$, they have achieved the two highest ER@5 scores. Upon detailed observation of the sub-figure depicting diffusion steps, we noticed that before reaching its peak at $T = 30$, the ER@5 scores increase rapidly while maintaining good image quality (with a slow increment in the FID score). Thus, based on the changes in ER@5 and FID for the diffusion steps, we contend that employing diffusion steps around 30 achieves a desirable ER@5 score while maintaining high image quality for the generated adversarial image.

\textbf{Impact of $\xi$.}
The guidance factor $\xi$ regulates the strength of guidance during the reverse process. In this paper, guidance is represented by the Mean Squared Error (MSE) loss between the original image and reversed images. Therefore, a larger $\xi$ will make the generated image more similar to the original image. By observing the bottom-right sub-figure of Figure~\ref{fig_hyperparams_fid}, the changes in ER@5 and FID exhibit highly similar trends. Based on these observations, a higher guidance scale corresponds to a better ER@5 score. Additionally, the overall FID scores are acceptable, as they are relatively low compared to other hyper-parameters, indicating good image quality. Thus, we argue that a relatively high guidance scale is essential to closely resemble the generated image to the original image and achieve better noise removal.

With cross-observations on the changes of ER@5 and FID among the four hyper-parameters (i.e., perturbation epsilon $\epsilon$, perturbation epochs $e$, diffusion steps $T$, and diffusion guide scale $\xi$), we claim that image quality emerges as a crucial implicit factor affecting not only the imperceptibility of IPDGI but also its effectiveness.

\subsection{Ablation Study}
\begin{figure}
    \centering
    \includegraphics[scale=0.6]{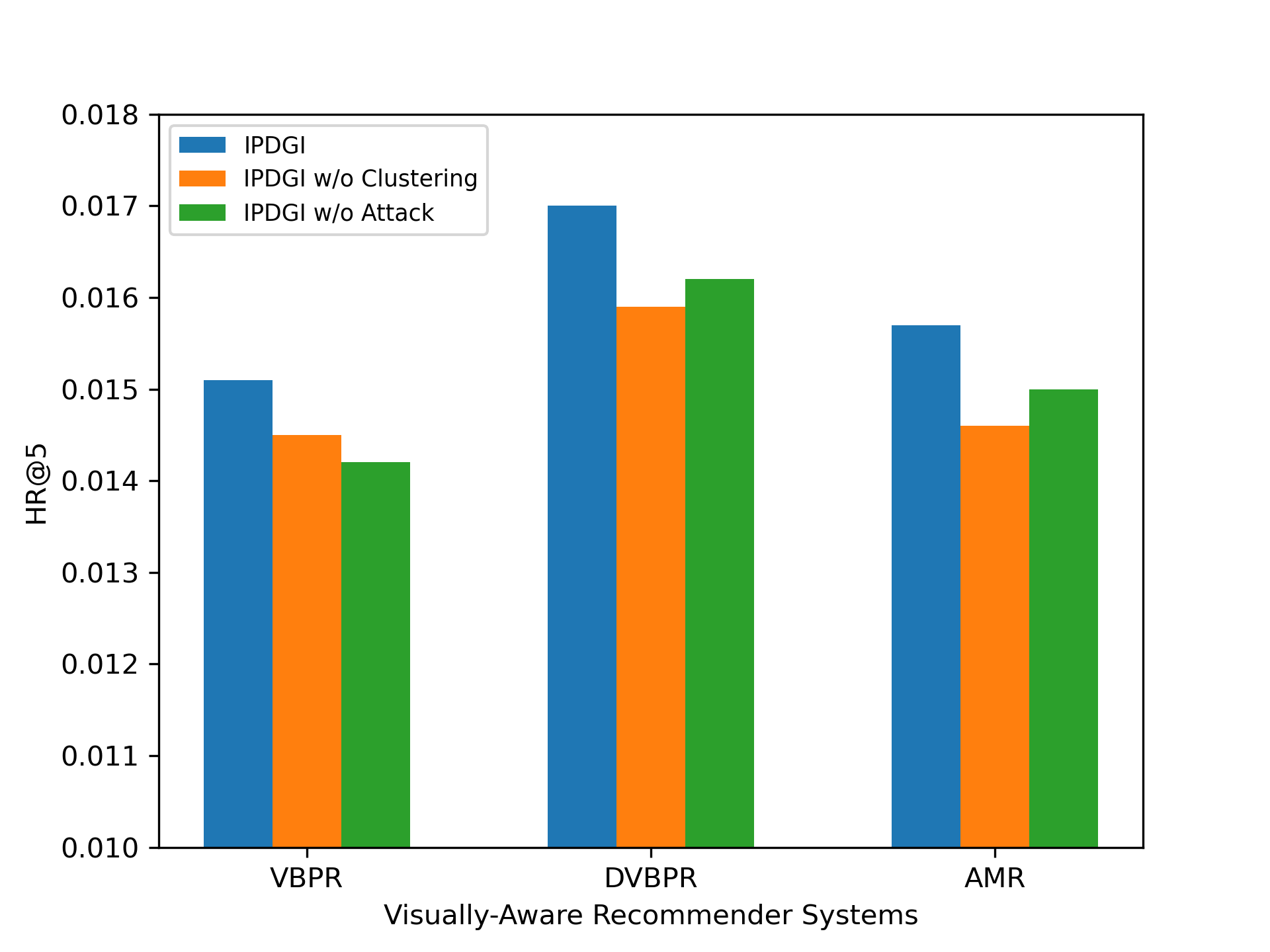}
    \caption{IPDGI Ablation Study}
    \label{fig_ablation}
\end{figure}

To validate the significance and necessity of each component of the IPDGI, we conducted an ablation study on three base visually-aware recommender systems using the Amazon Beauty dataset. The results are presented in Figure~\ref{fig_ablation}. 

For each recommender system, we compared exposure rate (ER@5) scores across three settings: IPDGI, IPDGI w/o Clustering, and IPDGI w/o Attack. In the ``IPDGI'' setting, recommender systems are evaluated with adversarial images generated by the fully functional IPDGI. In the ``IPDGI w/o Clustering'' setting, the image clustering model is disabled in the IPDGI, and the reference image is simply selected from the most popular item (i.e., the item with the most interactions in the dataset). Lastly, in the ``IPDGI w/o Attack'' setting, perturbations are not combined into the general Gaussian noise before the forward process of the diffusion. In other words, we only employ the base guided diffusion model (see Section~\ref{guided_diffusion}) to generate an image.

Based on the results depicted in Figure~\ref{fig_ablation}, it is evident that the fully-functional IPDGI achieves optimal ER@5 scores for all visually-aware recommender systems. Conversely, for the other two settings, a noteworthy decline in ER@5 performance is observed. The reduction in ER@5 observed in the ``IPDGI w/o Clustering'' setting highlights the significance of reference image selection. Additionally, the effect of the IPDGI may fluctuate depending on the dissimilarities or distances between clusters within the clustering model. Specifically, the greater the distance between the clusters in the clustering model, the more effective IPDGI becomes. The comparison between ``IPDGI'' and ``IPDGI w/o Attack'' implies the effectiveness of our perturbation generator. Additionally, an interesting observation is that ``IPDGI w/o Clustering'' even achieves poorer performance than without perturbation (i.e., ``IPDGI w/o Attack'') in some cases. This may be because the image of the most popular item may differ from some target images in terms of semantic content; therefore, using it as a reference image cannot guide the model to find the optimal perturbation.

\section{Discussion of Potential Defense Methods} \label{sec_defense}
Detecting adversarial images generated by IPDGI poses a significant challenge due to their high fidelity and imperceptibility to humans, particularly within visually-aware recommender systems. Unlike classifier-targeted attacks, IPDGI operates as a ranker-targeted attack, further complicating defense efforts. Potential defense methods primarily aim to diminish or eliminate the perturbations present in adversarial images. These methods generally fall into two categories: image compression and image reconstruction.

In the field of computer vision, image compression stands as a widely discussed defense strategy against adversarial images ~\cite{xu2017feature, guo2017countering, dziugaite2016study, das2018shield, jia2019comdefend}. This method involves preprocessing the input image (i.e., compress) to reduce adversarial perturbations before it feeds to the model. Importantly, image compression does not require retraining or modifying the model, rendering it a practical defense approach in real-world scenarios. However, image compression is unable to completely remove adversarial perturbations and may result in the loss of image information. Its effectiveness is also limited when faced with strong adversarial perturbations within adversarial images ~\cite{dai2020dipdefend, zhang2021defense}. Furthermore, in practice, the application of image compression would be applied to all images, potentially diminishing the utility of image information in visually-aware recommender systems ~\cite{liu2021adversarial}.

Similar to image compression, image reconstruction ~\cite{yuan2023manipulating1, dai2020dipdefend, zhang2021defense, song2017pixeldefend, yin2023securing, merra2023denoise} is another defense method that does not require retraining or modifying the model. Specifically, this approach involves generating a revised image through an image reconstruction network. The goal is to produce an image that appears identical to the original but without the adversarial perturbations. In contrast to image compression, image reconstruction tends to outperform in diminishing the perturbations present in adversarial images while preserving more image information.

\section{Conclusion and Future Work} \label{conclusion}
In this paper, we propose a novel attack specifically designed for visually-aware recommender systems, namely Item Promotion by Diffusion Guided Image (IPDGI). It adopts the diffusion model as the core framework for generating adversarial images to promote item rankings within the top-K ranker of the recommender model. To ensure the effectiveness of the IPDGI attack, we introduce an adversarial perturbation generator to produce optimized perturbations for the target item's image, effectively popularizing the item. Additionally, to maintain the imperceptibility of the IPDGI attack, we impose a conditional constraint at every time step of the reverse process of the diffusion model to preserve the visual consistency between the ultimate adversarial image and the original image. Extensive experiments conducted on two real-world datasets using three visually-aware recommender systems demonstrate the effectiveness and imperceptibility of our proposed attack. Furthermore, we conduct a hyper-parameters analysis and an ablation study to provide additional insights.
After highlighting the security hole of visually-aware recommender systems in this paper, in future research, we plan to explore the defense method and propose a more robust recommender system against the visual threat.

\begin{acks}
This work is supported by the Australian Research Council under the streams of Future Fellowship (Grant No. FT210100624), Discovery Early Career Researcher Award (Grants No. DE230101033 and No. DE200101465), Discovery Project (Grants No. DP240101108, and No. DP240101814), and Industrial Transformation Training Centre (Grant No. IC200100022).
\end{acks}

\bibliographystyle{ACM-Reference-Format}
\bibliography{sample-base}


\begin{thebibliography}{69}


\ifx \showCODEN    \undefined \def \showCODEN     #1{\unskip}     \fi
\ifx \showDOI      \undefined \def \showDOI       #1{#1}\fi
\ifx \showISBNx    \undefined \def \showISBNx     #1{\unskip}     \fi
\ifx \showISBNxiii \undefined \def \showISBNxiii  #1{\unskip}     \fi
\ifx \showISSN     \undefined \def \showISSN      #1{\unskip}     \fi
\ifx \showLCCN     \undefined \def \showLCCN      #1{\unskip}     \fi
\ifx \shownote     \undefined \def \shownote      #1{#1}          \fi
\ifx \showarticletitle \undefined \def \showarticletitle #1{#1}   \fi
\ifx \showURL      \undefined \def \showURL       {\relax}        \fi
\providecommand\bibfield[2]{#2}
\providecommand\bibinfo[2]{#2}
\providecommand\natexlab[1]{#1}
\providecommand\showeprint[2][]{arXiv:#2}

\bibitem[\protect\citeauthoryear{Bracher, Heinz, and Vollgraf}{Bracher
  et~al\mbox{.}}{2016}]%
        {bracher2016fashion}
\bibfield{author}{\bibinfo{person}{Christian Bracher},
  \bibinfo{person}{Sebastian Heinz}, {and} \bibinfo{person}{Roland Vollgraf}.}
  \bibinfo{year}{2016}\natexlab{}.
\newblock \showarticletitle{Fashion DNA: merging content and sales data for
  recommendation and article mapping}.
\newblock \bibinfo{journal}{\emph{arXiv preprint arXiv:1609.02489}}
  (\bibinfo{year}{2016}).
\newblock


\bibitem[\protect\citeauthoryear{Chen, Zhang, He, Nie, Liu, and Chua}{Chen
  et~al\mbox{.}}{2017}]%
        {chen2017attentive}
\bibfield{author}{\bibinfo{person}{Jingyuan Chen}, \bibinfo{person}{Hanwang
  Zhang}, \bibinfo{person}{Xiangnan He}, \bibinfo{person}{Liqiang Nie},
  \bibinfo{person}{Wei Liu}, {and} \bibinfo{person}{Tat-Seng Chua}.}
  \bibinfo{year}{2017}\natexlab{}.
\newblock \showarticletitle{Attentive collaborative filtering: Multimedia
  recommendation with item-and component-level attention}. In
  \bibinfo{booktitle}{\emph{Proceedings of the 40th International ACM SIGIR
  conference on Research and Development in Information Retrieval}}.
  \bibinfo{pages}{335--344}.
\newblock


\bibitem[\protect\citeauthoryear{Chen, Yin, Chen, Wu, Wang, Zhou, and Li}{Chen
  et~al\mbox{.}}{2018}]%
        {chen2018tada}
\bibfield{author}{\bibinfo{person}{Tong Chen}, \bibinfo{person}{Hongzhi Yin},
  \bibinfo{person}{Hongxu Chen}, \bibinfo{person}{Lin Wu}, \bibinfo{person}{Hao
  Wang}, \bibinfo{person}{Xiaofang Zhou}, {and} \bibinfo{person}{Xue Li}.}
  \bibinfo{year}{2018}\natexlab{}.
\newblock \showarticletitle{Tada: trend alignment with dual-attention
  multi-task recurrent neural networks for sales prediction}. In
  \bibinfo{booktitle}{\emph{2018 IEEE international conference on data mining
  (ICDM)}}. IEEE, \bibinfo{pages}{49--58}.
\newblock


\bibitem[\protect\citeauthoryear{Cheng, Pan, Zhang, Ni, Sun, and Yuan}{Cheng
  et~al\mbox{.}}{2023}]%
        {cheng2023image}
\bibfield{author}{\bibinfo{person}{Yu Cheng}, \bibinfo{person}{Yunzhu Pan},
  \bibinfo{person}{Jiaqi Zhang}, \bibinfo{person}{Yongxin Ni},
  \bibinfo{person}{Aixin Sun}, {and} \bibinfo{person}{Fajie Yuan}.}
  \bibinfo{year}{2023}\natexlab{}.
\newblock \showarticletitle{An Image Dataset for Benchmarking Recommender
  Systems with Raw Pixels}.
\newblock \bibinfo{journal}{\emph{arXiv preprint arXiv:2309.06789}}
  (\bibinfo{year}{2023}).
\newblock


\bibitem[\protect\citeauthoryear{Cohen, Sar~Shalom, Jannach, and Amir}{Cohen
  et~al\mbox{.}}{2021}]%
        {cohen2021black}
\bibfield{author}{\bibinfo{person}{Rami Cohen}, \bibinfo{person}{Oren
  Sar~Shalom}, \bibinfo{person}{Dietmar Jannach}, {and}
  \bibinfo{person}{Amihood Amir}.} \bibinfo{year}{2021}\natexlab{}.
\newblock \showarticletitle{A black-box attack model for visually-aware
  recommender systems}. In \bibinfo{booktitle}{\emph{Proceedings of the 14th
  ACM International Conference on Web Search and Data Mining}}.
  \bibinfo{pages}{94--102}.
\newblock


\bibitem[\protect\citeauthoryear{Dai, Feng, Wu, Chen, Lu, Jiang, and Xia}{Dai
  et~al\mbox{.}}{2020}]%
        {dai2020dipdefend}
\bibfield{author}{\bibinfo{person}{Tao Dai}, \bibinfo{person}{Yan Feng},
  \bibinfo{person}{Dongxian Wu}, \bibinfo{person}{Bin Chen},
  \bibinfo{person}{Jian Lu}, \bibinfo{person}{Yong Jiang}, {and}
  \bibinfo{person}{Shu-Tao Xia}.} \bibinfo{year}{2020}\natexlab{}.
\newblock \showarticletitle{Dipdefend: Deep image prior driven defense against
  adversarial examples}. In \bibinfo{booktitle}{\emph{Proceedings of the 28th
  ACM International Conference on Multimedia}}. \bibinfo{pages}{1404--1412}.
\newblock


\bibitem[\protect\citeauthoryear{Das, Shanbhogue, Chen, Hohman, Li, Chen,
  Kounavis, and Chau}{Das et~al\mbox{.}}{2018}]%
        {das2018shield}
\bibfield{author}{\bibinfo{person}{Nilaksh Das}, \bibinfo{person}{Madhuri
  Shanbhogue}, \bibinfo{person}{Shang-Tse Chen}, \bibinfo{person}{Fred Hohman},
  \bibinfo{person}{Siwei Li}, \bibinfo{person}{Li Chen},
  \bibinfo{person}{Michael~E Kounavis}, {and} \bibinfo{person}{Duen~Horng
  Chau}.} \bibinfo{year}{2018}\natexlab{}.
\newblock \showarticletitle{Shield: Fast, practical defense and vaccination for
  deep learning using jpeg compression}. In
  \bibinfo{booktitle}{\emph{Proceedings of the 24th ACM SIGKDD International
  Conference on Knowledge Discovery \& Data Mining}}.
  \bibinfo{pages}{196--204}.
\newblock


\bibitem[\protect\citeauthoryear{Dhariwal and Nichol}{Dhariwal and
  Nichol}{2021}]%
        {dhariwal2021diffusion}
\bibfield{author}{\bibinfo{person}{Prafulla Dhariwal} {and}
  \bibinfo{person}{Alexander Nichol}.} \bibinfo{year}{2021}\natexlab{}.
\newblock \showarticletitle{Diffusion models beat gans on image synthesis}.
\newblock \bibinfo{journal}{\emph{Advances in neural information processing
  systems}}  \bibinfo{volume}{34} (\bibinfo{year}{2021}),
  \bibinfo{pages}{8780--8794}.
\newblock


\bibitem[\protect\citeauthoryear{Di~Noia, Malitesta, and Merra}{Di~Noia
  et~al\mbox{.}}{2020}]%
        {di2020taamr}
\bibfield{author}{\bibinfo{person}{Tommaso Di~Noia}, \bibinfo{person}{Daniele
  Malitesta}, {and} \bibinfo{person}{Felice~Antonio Merra}.}
  \bibinfo{year}{2020}\natexlab{}.
\newblock \showarticletitle{Taamr: Targeted adversarial attack against
  multimedia recommender systems}. In \bibinfo{booktitle}{\emph{2020 50th
  Annual IEEE/IFIP international conference on dependable systems and networks
  workshops (DSN-W)}}. IEEE, \bibinfo{pages}{1--8}.
\newblock


\bibitem[\protect\citeauthoryear{Du, Yuan, Huang, Zhao, and Zhou}{Du
  et~al\mbox{.}}{2023}]%
        {du2023sequential}
\bibfield{author}{\bibinfo{person}{Hanwen Du}, \bibinfo{person}{Huanhuan Yuan},
  \bibinfo{person}{Zhen Huang}, \bibinfo{person}{Pengpeng Zhao}, {and}
  \bibinfo{person}{Xiaofang Zhou}.} \bibinfo{year}{2023}\natexlab{}.
\newblock \showarticletitle{Sequential Recommendation with Diffusion Models}.
\newblock \bibinfo{journal}{\emph{arXiv preprint arXiv:2304.04541}}
  (\bibinfo{year}{2023}).
\newblock


\bibitem[\protect\citeauthoryear{Dziugaite, Ghahramani, and Roy}{Dziugaite
  et~al\mbox{.}}{2016}]%
        {dziugaite2016study}
\bibfield{author}{\bibinfo{person}{Gintare~Karolina Dziugaite},
  \bibinfo{person}{Zoubin Ghahramani}, {and} \bibinfo{person}{Daniel~M Roy}.}
  \bibinfo{year}{2016}\natexlab{}.
\newblock \showarticletitle{A study of the effect of jpg compression on
  adversarial images}.
\newblock \bibinfo{journal}{\emph{arXiv preprint arXiv:1608.00853}}
  (\bibinfo{year}{2016}).
\newblock


\bibitem[\protect\citeauthoryear{Elsayed, Shankar, Cheung, Papernot, Kurakin,
  Goodfellow, and Sohl-Dickstein}{Elsayed et~al\mbox{.}}{2018}]%
        {elsayed2018adversarial}
\bibfield{author}{\bibinfo{person}{Gamaleldin Elsayed}, \bibinfo{person}{Shreya
  Shankar}, \bibinfo{person}{Brian Cheung}, \bibinfo{person}{Nicolas Papernot},
  \bibinfo{person}{Alexey Kurakin}, \bibinfo{person}{Ian Goodfellow}, {and}
  \bibinfo{person}{Jascha Sohl-Dickstein}.} \bibinfo{year}{2018}\natexlab{}.
\newblock \showarticletitle{Adversarial examples that fool both computer vision
  and time-limited humans}.
\newblock \bibinfo{journal}{\emph{Advances in neural information processing
  systems}}  \bibinfo{volume}{31} (\bibinfo{year}{2018}).
\newblock


\bibitem[\protect\citeauthoryear{Goodfellow, Pouget-Abadie, Mirza, Xu,
  Warde-Farley, Ozair, Courville, and Bengio}{Goodfellow et~al\mbox{.}}{2020}]%
        {goodfellow2020generative}
\bibfield{author}{\bibinfo{person}{Ian Goodfellow}, \bibinfo{person}{Jean
  Pouget-Abadie}, \bibinfo{person}{Mehdi Mirza}, \bibinfo{person}{Bing Xu},
  \bibinfo{person}{David Warde-Farley}, \bibinfo{person}{Sherjil Ozair},
  \bibinfo{person}{Aaron Courville}, {and} \bibinfo{person}{Yoshua Bengio}.}
  \bibinfo{year}{2020}\natexlab{}.
\newblock \showarticletitle{Generative adversarial networks}.
\newblock \bibinfo{journal}{\emph{Commun. ACM}} \bibinfo{volume}{63},
  \bibinfo{number}{11} (\bibinfo{year}{2020}), \bibinfo{pages}{139--144}.
\newblock


\bibitem[\protect\citeauthoryear{Goodfellow, Shlens, and Szegedy}{Goodfellow
  et~al\mbox{.}}{2014}]%
        {goodfellow2014explaining}
\bibfield{author}{\bibinfo{person}{Ian~J Goodfellow}, \bibinfo{person}{Jonathon
  Shlens}, {and} \bibinfo{person}{Christian Szegedy}.}
  \bibinfo{year}{2014}\natexlab{}.
\newblock \showarticletitle{Explaining and harnessing adversarial examples}.
\newblock \bibinfo{journal}{\emph{arXiv preprint arXiv:1412.6572}}
  (\bibinfo{year}{2014}).
\newblock


\bibitem[\protect\citeauthoryear{Guo, Rana, Cisse, and Van Der~Maaten}{Guo
  et~al\mbox{.}}{2017}]%
        {guo2017countering}
\bibfield{author}{\bibinfo{person}{Chuan Guo}, \bibinfo{person}{Mayank Rana},
  \bibinfo{person}{Moustapha Cisse}, {and} \bibinfo{person}{Laurens Van
  Der~Maaten}.} \bibinfo{year}{2017}\natexlab{}.
\newblock \showarticletitle{Countering adversarial images using input
  transformations}.
\newblock \bibinfo{journal}{\emph{arXiv preprint arXiv:1711.00117}}
  (\bibinfo{year}{2017}).
\newblock


\bibitem[\protect\citeauthoryear{He, Zhang, Ren, and Sun}{He
  et~al\mbox{.}}{2016}]%
        {he2016deep}
\bibfield{author}{\bibinfo{person}{Kaiming He}, \bibinfo{person}{Xiangyu
  Zhang}, \bibinfo{person}{Shaoqing Ren}, {and} \bibinfo{person}{Jian Sun}.}
  \bibinfo{year}{2016}\natexlab{}.
\newblock \showarticletitle{Deep residual learning for image recognition}. In
  \bibinfo{booktitle}{\emph{Proceedings of the IEEE conference on computer
  vision and pattern recognition}}. \bibinfo{pages}{770--778}.
\newblock


\bibitem[\protect\citeauthoryear{He and McAuley}{He and McAuley}{2016}]%
        {he2016vbpr}
\bibfield{author}{\bibinfo{person}{Ruining He} {and} \bibinfo{person}{Julian
  McAuley}.} \bibinfo{year}{2016}\natexlab{}.
\newblock \showarticletitle{VBPR: visual bayesian personalized ranking from
  implicit feedback}. In \bibinfo{booktitle}{\emph{Proceedings of the AAAI
  conference on artificial intelligence}}, Vol.~\bibinfo{volume}{30}.
\newblock


\bibitem[\protect\citeauthoryear{He, Liao, Zhang, Nie, Hu, and Chua}{He
  et~al\mbox{.}}{2017}]%
        {he2017neural}
\bibfield{author}{\bibinfo{person}{Xiangnan He}, \bibinfo{person}{Lizi Liao},
  \bibinfo{person}{Hanwang Zhang}, \bibinfo{person}{Liqiang Nie},
  \bibinfo{person}{Xia Hu}, {and} \bibinfo{person}{Tat-Seng Chua}.}
  \bibinfo{year}{2017}\natexlab{}.
\newblock \showarticletitle{Neural collaborative filtering}. In
  \bibinfo{booktitle}{\emph{Proceedings of the 26th international conference on
  world wide web}}. \bibinfo{pages}{173--182}.
\newblock


\bibitem[\protect\citeauthoryear{Heusel, Ramsauer, Unterthiner, Nessler, and
  Hochreiter}{Heusel et~al\mbox{.}}{2017}]%
        {heusel2017gans}
\bibfield{author}{\bibinfo{person}{Martin Heusel}, \bibinfo{person}{Hubert
  Ramsauer}, \bibinfo{person}{Thomas Unterthiner}, \bibinfo{person}{Bernhard
  Nessler}, {and} \bibinfo{person}{Sepp Hochreiter}.}
  \bibinfo{year}{2017}\natexlab{}.
\newblock \showarticletitle{Gans trained by a two time-scale update rule
  converge to a local nash equilibrium}.
\newblock \bibinfo{journal}{\emph{Advances in neural information processing
  systems}}  \bibinfo{volume}{30} (\bibinfo{year}{2017}).
\newblock


\bibitem[\protect\citeauthoryear{Ho, Jain, and Abbeel}{Ho
  et~al\mbox{.}}{2020}]%
        {ho2020denoising}
\bibfield{author}{\bibinfo{person}{Jonathan Ho}, \bibinfo{person}{Ajay Jain},
  {and} \bibinfo{person}{Pieter Abbeel}.} \bibinfo{year}{2020}\natexlab{}.
\newblock \showarticletitle{Denoising diffusion probabilistic models}.
\newblock \bibinfo{journal}{\emph{Advances in neural information processing
  systems}}  \bibinfo{volume}{33} (\bibinfo{year}{2020}),
  \bibinfo{pages}{6840--6851}.
\newblock


\bibitem[\protect\citeauthoryear{Jagadeesh, Piramuthu, Bhardwaj, Di, and
  Sundaresan}{Jagadeesh et~al\mbox{.}}{2014}]%
        {jagadeesh2014large}
\bibfield{author}{\bibinfo{person}{Vignesh Jagadeesh},
  \bibinfo{person}{Robinson Piramuthu}, \bibinfo{person}{Anurag Bhardwaj},
  \bibinfo{person}{Wei Di}, {and} \bibinfo{person}{Neel Sundaresan}.}
  \bibinfo{year}{2014}\natexlab{}.
\newblock \showarticletitle{Large scale visual recommendations from street
  fashion images}. In \bibinfo{booktitle}{\emph{Proceedings of the 20th ACM
  SIGKDD international conference on Knowledge discovery and data mining}}.
  \bibinfo{pages}{1925--1934}.
\newblock


\bibitem[\protect\citeauthoryear{Jia, Wei, Cao, and Foroosh}{Jia
  et~al\mbox{.}}{2019}]%
        {jia2019comdefend}
\bibfield{author}{\bibinfo{person}{Xiaojun Jia}, \bibinfo{person}{Xingxing
  Wei}, \bibinfo{person}{Xiaochun Cao}, {and} \bibinfo{person}{Hassan
  Foroosh}.} \bibinfo{year}{2019}\natexlab{}.
\newblock \showarticletitle{Comdefend: An efficient image compression model to
  defend adversarial examples}. In \bibinfo{booktitle}{\emph{Proceedings of the
  IEEE/CVF conference on computer vision and pattern recognition}}.
  \bibinfo{pages}{6084--6092}.
\newblock


\bibitem[\protect\citeauthoryear{Kalantidis, Kennedy, and Li}{Kalantidis
  et~al\mbox{.}}{2013}]%
        {kalantidis2013getting}
\bibfield{author}{\bibinfo{person}{Yannis Kalantidis}, \bibinfo{person}{Lyndon
  Kennedy}, {and} \bibinfo{person}{Li-Jia Li}.}
  \bibinfo{year}{2013}\natexlab{}.
\newblock \showarticletitle{Getting the look: clothing recognition and
  segmentation for automatic product suggestions in everyday photos}. In
  \bibinfo{booktitle}{\emph{Proceedings of the 3rd ACM conference on
  International conference on multimedia retrieval}}.
  \bibinfo{pages}{105--112}.
\newblock


\bibitem[\protect\citeauthoryear{Kang, Fang, Wang, and McAuley}{Kang
  et~al\mbox{.}}{2017}]%
        {kang2017visually}
\bibfield{author}{\bibinfo{person}{Wang-Cheng Kang}, \bibinfo{person}{Chen
  Fang}, \bibinfo{person}{Zhaowen Wang}, {and} \bibinfo{person}{Julian
  McAuley}.} \bibinfo{year}{2017}\natexlab{}.
\newblock \showarticletitle{Visually-aware fashion recommendation and design
  with generative image models}. In \bibinfo{booktitle}{\emph{2017 IEEE
  international conference on data mining (ICDM)}}. IEEE,
  \bibinfo{pages}{207--216}.
\newblock


\bibitem[\protect\citeauthoryear{Kingma and Ba}{Kingma and Ba}{2014}]%
        {kingma2014adam}
\bibfield{author}{\bibinfo{person}{Diederik~P Kingma} {and}
  \bibinfo{person}{Jimmy Ba}.} \bibinfo{year}{2014}\natexlab{}.
\newblock \showarticletitle{Adam: A method for stochastic optimization}.
\newblock \bibinfo{journal}{\emph{arXiv preprint arXiv:1412.6980}}
  (\bibinfo{year}{2014}).
\newblock


\bibitem[\protect\citeauthoryear{Kingma and Welling}{Kingma and
  Welling}{2013}]%
        {kingma2013auto}
\bibfield{author}{\bibinfo{person}{Diederik~P Kingma} {and}
  \bibinfo{person}{Max Welling}.} \bibinfo{year}{2013}\natexlab{}.
\newblock \showarticletitle{Auto-encoding variational bayes}.
\newblock \bibinfo{journal}{\emph{arXiv preprint arXiv:1312.6114}}
  (\bibinfo{year}{2013}).
\newblock


\bibitem[\protect\citeauthoryear{Kurakin, Goodfellow, and Bengio}{Kurakin
  et~al\mbox{.}}{2018}]%
        {kurakin2018adversarial}
\bibfield{author}{\bibinfo{person}{Alexey Kurakin}, \bibinfo{person}{Ian~J
  Goodfellow}, {and} \bibinfo{person}{Samy Bengio}.}
  \bibinfo{year}{2018}\natexlab{}.
\newblock \showarticletitle{Adversarial examples in the physical world}.
\newblock In \bibinfo{booktitle}{\emph{Artificial intelligence safety and
  security}}. \bibinfo{publisher}{Chapman and Hall/CRC},
  \bibinfo{pages}{99--112}.
\newblock


\bibitem[\protect\citeauthoryear{Lei, Liu, Li, Zha, and Li}{Lei
  et~al\mbox{.}}{2016}]%
        {lei2016comparative}
\bibfield{author}{\bibinfo{person}{Chenyi Lei}, \bibinfo{person}{Dong Liu},
  \bibinfo{person}{Weiping Li}, \bibinfo{person}{Zheng-Jun Zha}, {and}
  \bibinfo{person}{Houqiang Li}.} \bibinfo{year}{2016}\natexlab{}.
\newblock \showarticletitle{Comparative deep learning of hybrid representations
  for image recommendations}. In \bibinfo{booktitle}{\emph{Proceedings of the
  IEEE conference on computer vision and pattern recognition}}.
  \bibinfo{pages}{2545--2553}.
\newblock


\bibitem[\protect\citeauthoryear{Li, Chen, Zhang, and Yin}{Li
  et~al\mbox{.}}{2021}]%
        {li2021lightweight}
\bibfield{author}{\bibinfo{person}{Yang Li}, \bibinfo{person}{Tong Chen},
  \bibinfo{person}{Peng-Fei Zhang}, {and} \bibinfo{person}{Hongzhi Yin}.}
  \bibinfo{year}{2021}\natexlab{}.
\newblock \showarticletitle{Lightweight self-attentive sequential
  recommendation}. In \bibinfo{booktitle}{\emph{Proceedings of the 30th ACM
  International Conference on Information \& Knowledge Management}}.
  \bibinfo{pages}{967--977}.
\newblock


\bibitem[\protect\citeauthoryear{Li, Sun, and Li}{Li et~al\mbox{.}}{2023}]%
        {li2023diffurec}
\bibfield{author}{\bibinfo{person}{Zihao Li}, \bibinfo{person}{Aixin Sun},
  {and} \bibinfo{person}{Chenliang Li}.} \bibinfo{year}{2023}\natexlab{}.
\newblock \showarticletitle{DiffuRec: A Diffusion Model for Sequential
  Recommendation}.
\newblock \bibinfo{journal}{\emph{arXiv preprint arXiv:2304.00686}}
  (\bibinfo{year}{2023}).
\newblock


\bibitem[\protect\citeauthoryear{Liu and Larson}{Liu and Larson}{2021}]%
        {liu2021adversarial}
\bibfield{author}{\bibinfo{person}{Zhuoran Liu} {and} \bibinfo{person}{Martha
  Larson}.} \bibinfo{year}{2021}\natexlab{}.
\newblock \showarticletitle{Adversarial item promotion: Vulnerabilities at the
  core of top-n recommenders that use images to address cold start}. In
  \bibinfo{booktitle}{\emph{Proceedings of the Web Conference 2021}}.
  \bibinfo{pages}{3590--3602}.
\newblock


\bibitem[\protect\citeauthoryear{Long, Gao, Xu, and Zhou}{Long
  et~al\mbox{.}}{2022}]%
        {long2022survey}
\bibfield{author}{\bibinfo{person}{Teng Long}, \bibinfo{person}{Qi Gao},
  \bibinfo{person}{Lili Xu}, {and} \bibinfo{person}{Zhangbing Zhou}.}
  \bibinfo{year}{2022}\natexlab{}.
\newblock \showarticletitle{A survey on adversarial attacks in computer vision:
  Taxonomy, visualization and future directions}.
\newblock \bibinfo{journal}{\emph{Computers \& Security}}
  (\bibinfo{year}{2022}), \bibinfo{pages}{102847}.
\newblock


\bibitem[\protect\citeauthoryear{Magnusson, Andersen, Jonasson, and
  Vehtari}{Magnusson et~al\mbox{.}}{2019}]%
        {magnusson2019bayesian}
\bibfield{author}{\bibinfo{person}{M{\aa}ns Magnusson},
  \bibinfo{person}{Michael Andersen}, \bibinfo{person}{Johan Jonasson}, {and}
  \bibinfo{person}{Aki Vehtari}.} \bibinfo{year}{2019}\natexlab{}.
\newblock \showarticletitle{Bayesian leave-one-out cross-validation for large
  data}. In \bibinfo{booktitle}{\emph{International Conference on Machine
  Learning}}. PMLR, \bibinfo{pages}{4244--4253}.
\newblock


\bibitem[\protect\citeauthoryear{McAuley, Targett, Shi, and Van
  Den~Hengel}{McAuley et~al\mbox{.}}{2015}]%
        {mcauley2015image}
\bibfield{author}{\bibinfo{person}{Julian McAuley},
  \bibinfo{person}{Christopher Targett}, \bibinfo{person}{Qinfeng Shi}, {and}
  \bibinfo{person}{Anton Van Den~Hengel}.} \bibinfo{year}{2015}\natexlab{}.
\newblock \showarticletitle{Image-based recommendations on styles and
  substitutes}. In \bibinfo{booktitle}{\emph{Proceedings of the 38th
  international ACM SIGIR conference on research and development in information
  retrieval}}. \bibinfo{pages}{43--52}.
\newblock


\bibitem[\protect\citeauthoryear{Merra, Anelli, Di~Noia, Malitesta, and
  Mancino}{Merra et~al\mbox{.}}{2023}]%
        {merra2023denoise}
\bibfield{author}{\bibinfo{person}{Felice~Antonio Merra},
  \bibinfo{person}{Vito~Walter Anelli}, \bibinfo{person}{Tommaso Di~Noia},
  \bibinfo{person}{Daniele Malitesta}, {and} \bibinfo{person}{Alberto
  Carlo~Maria Mancino}.} \bibinfo{year}{2023}\natexlab{}.
\newblock \showarticletitle{Denoise to Protect: A Method to Robustify Visual
  Recommenders from Adversaries}. In \bibinfo{booktitle}{\emph{Proceedings of
  the 46th International ACM SIGIR Conference on Research and Development in
  Information Retrieval}}. \bibinfo{pages}{1924--1928}.
\newblock


\bibitem[\protect\citeauthoryear{Neve and McConville}{Neve and
  McConville}{2020}]%
        {neve2020imrec}
\bibfield{author}{\bibinfo{person}{James Neve} {and} \bibinfo{person}{Ryan
  McConville}.} \bibinfo{year}{2020}\natexlab{}.
\newblock \showarticletitle{ImRec: Learning reciprocal preferences using
  images}. In \bibinfo{booktitle}{\emph{Proceedings of the 14th ACM Conference
  on Recommender Systems}}. \bibinfo{pages}{170--179}.
\newblock


\bibitem[\protect\citeauthoryear{Nguyen, Huynh, Nguyen, Liew, Yin, and
  Nguyen}{Nguyen et~al\mbox{.}}{2022}]%
        {nguyen2022survey}
\bibfield{author}{\bibinfo{person}{Thanh~Tam Nguyen},
  \bibinfo{person}{Thanh~Trung Huynh}, \bibinfo{person}{Phi~Le Nguyen},
  \bibinfo{person}{Alan Wee-Chung Liew}, \bibinfo{person}{Hongzhi Yin}, {and}
  \bibinfo{person}{Quoc Viet~Hung Nguyen}.} \bibinfo{year}{2022}\natexlab{}.
\newblock \showarticletitle{A survey of machine unlearning}.
\newblock \bibinfo{journal}{\emph{arXiv preprint arXiv:2209.02299}}
  (\bibinfo{year}{2022}).
\newblock


\bibitem[\protect\citeauthoryear{Nguyen, Nguyen, Nguyen, Huynh, Nguyen,
  Weidlich, and Yin}{Nguyen et~al\mbox{.}}{2024}]%
        {nguyen2024manipulating}
\bibfield{author}{\bibinfo{person}{Thanh~Toan Nguyen}, \bibinfo{person}{Quoc
  Viet~Hung Nguyen}, \bibinfo{person}{Thanh~Tam Nguyen},
  \bibinfo{person}{Thanh~Trung Huynh}, \bibinfo{person}{Thanh~Thi Nguyen},
  \bibinfo{person}{Matthias Weidlich}, {and} \bibinfo{person}{Hongzhi Yin}.}
  \bibinfo{year}{2024}\natexlab{}.
\newblock \showarticletitle{Manipulating Recommender Systems: A Survey of
  Poisoning Attacks and Countermeasures}.
\newblock \bibinfo{journal}{\emph{arXiv preprint arXiv:2404.14942}}
  (\bibinfo{year}{2024}).
\newblock


\bibitem[\protect\citeauthoryear{Paszke, Gross, Massa, Lerer, Bradbury, Chanan,
  Killeen, Lin, Gimelshein, Antiga, et~al\mbox{.}}{Paszke
  et~al\mbox{.}}{2019}]%
        {paszke2019pytorch}
\bibfield{author}{\bibinfo{person}{Adam Paszke}, \bibinfo{person}{Sam Gross},
  \bibinfo{person}{Francisco Massa}, \bibinfo{person}{Adam Lerer},
  \bibinfo{person}{James Bradbury}, \bibinfo{person}{Gregory Chanan},
  \bibinfo{person}{Trevor Killeen}, \bibinfo{person}{Zeming Lin},
  \bibinfo{person}{Natalia Gimelshein}, \bibinfo{person}{Luca Antiga},
  {et~al\mbox{.}}} \bibinfo{year}{2019}\natexlab{}.
\newblock \showarticletitle{Pytorch: An imperative style, high-performance deep
  learning library}.
\newblock \bibinfo{journal}{\emph{Advances in neural information processing
  systems}}  \bibinfo{volume}{32} (\bibinfo{year}{2019}).
\newblock


\bibitem[\protect\citeauthoryear{Qiu, Li, Huang, and Yin}{Qiu
  et~al\mbox{.}}{2019}]%
        {qiu2019rethinking}
\bibfield{author}{\bibinfo{person}{Ruihong Qiu}, \bibinfo{person}{Jingjing Li},
  \bibinfo{person}{Zi Huang}, {and} \bibinfo{person}{Hongzhi Yin}.}
  \bibinfo{year}{2019}\natexlab{}.
\newblock \showarticletitle{Rethinking the item order in session-based
  recommendation with graph neural networks}. In
  \bibinfo{booktitle}{\emph{Proceedings of the 28th ACM international
  conference on information and knowledge management}}.
  \bibinfo{pages}{579--588}.
\newblock


\bibitem[\protect\citeauthoryear{Qiu, Yin, Huang, and Chen}{Qiu
  et~al\mbox{.}}{2020}]%
        {qiu2020gag}
\bibfield{author}{\bibinfo{person}{Ruihong Qiu}, \bibinfo{person}{Hongzhi Yin},
  \bibinfo{person}{Zi Huang}, {and} \bibinfo{person}{Tong Chen}.}
  \bibinfo{year}{2020}\natexlab{}.
\newblock \showarticletitle{Gag: Global attributed graph neural network for
  streaming session-based recommendation}. In
  \bibinfo{booktitle}{\emph{Proceedings of the 43rd International ACM SIGIR
  Conference on Research and Development in Information Retrieval}}.
  \bibinfo{pages}{669--678}.
\newblock


\bibitem[\protect\citeauthoryear{Qu, Tang, Zheng, Nguyen, Huang, Shi, and
  Yin}{Qu et~al\mbox{.}}{2023}]%
        {qu2023semi}
\bibfield{author}{\bibinfo{person}{Liang Qu}, \bibinfo{person}{Ningzhi Tang},
  \bibinfo{person}{Ruiqi Zheng}, \bibinfo{person}{Quoc Viet~Hung Nguyen},
  \bibinfo{person}{Zi Huang}, \bibinfo{person}{Yuhui Shi}, {and}
  \bibinfo{person}{Hongzhi Yin}.} \bibinfo{year}{2023}\natexlab{}.
\newblock \showarticletitle{Semi-decentralized Federated Ego Graph Learning for
  Recommendation}. In \bibinfo{booktitle}{\emph{Proceedings of the ACM Web
  Conference 2023}}. \bibinfo{pages}{339--348}.
\newblock


\bibitem[\protect\citeauthoryear{Qu, Zhu, Zheng, Shi, and Yin}{Qu
  et~al\mbox{.}}{2021}]%
        {qu2021imgagn}
\bibfield{author}{\bibinfo{person}{Liang Qu}, \bibinfo{person}{Huaisheng Zhu},
  \bibinfo{person}{Ruiqi Zheng}, \bibinfo{person}{Yuhui Shi}, {and}
  \bibinfo{person}{Hongzhi Yin}.} \bibinfo{year}{2021}\natexlab{}.
\newblock \showarticletitle{Imgagn: Imbalanced network embedding via generative
  adversarial graph networks}. In \bibinfo{booktitle}{\emph{Proceedings of the
  27th ACM SIGKDD Conference on Knowledge Discovery \& Data Mining}}.
  \bibinfo{pages}{1390--1398}.
\newblock


\bibitem[\protect\citeauthoryear{Rendle, Freudenthaler, Gantner, and
  Schmidt-Thieme}{Rendle et~al\mbox{.}}{2012}]%
        {rendle2012bpr}
\bibfield{author}{\bibinfo{person}{Steffen Rendle}, \bibinfo{person}{Christoph
  Freudenthaler}, \bibinfo{person}{Zeno Gantner}, {and} \bibinfo{person}{Lars
  Schmidt-Thieme}.} \bibinfo{year}{2012}\natexlab{}.
\newblock \showarticletitle{BPR: Bayesian personalized ranking from implicit
  feedback}.
\newblock \bibinfo{journal}{\emph{arXiv preprint arXiv:1205.2618}}
  (\bibinfo{year}{2012}).
\newblock


\bibitem[\protect\citeauthoryear{Russakovsky, Deng, Su, Krause, Satheesh, Ma,
  Huang, Karpathy, Khosla, Bernstein, et~al\mbox{.}}{Russakovsky
  et~al\mbox{.}}{2015}]%
        {russakovsky2015imagenet}
\bibfield{author}{\bibinfo{person}{Olga Russakovsky}, \bibinfo{person}{Jia
  Deng}, \bibinfo{person}{Hao Su}, \bibinfo{person}{Jonathan Krause},
  \bibinfo{person}{Sanjeev Satheesh}, \bibinfo{person}{Sean Ma},
  \bibinfo{person}{Zhiheng Huang}, \bibinfo{person}{Andrej Karpathy},
  \bibinfo{person}{Aditya Khosla}, \bibinfo{person}{Michael Bernstein},
  {et~al\mbox{.}}} \bibinfo{year}{2015}\natexlab{}.
\newblock \showarticletitle{Imagenet large scale visual recognition challenge}.
\newblock \bibinfo{journal}{\emph{International journal of computer vision}}
  \bibinfo{volume}{115} (\bibinfo{year}{2015}), \bibinfo{pages}{211--252}.
\newblock


\bibitem[\protect\citeauthoryear{Schein, Popescul, Ungar, and Pennock}{Schein
  et~al\mbox{.}}{2002}]%
        {schein2002methods}
\bibfield{author}{\bibinfo{person}{Andrew~I Schein},
  \bibinfo{person}{Alexandrin Popescul}, \bibinfo{person}{Lyle~H Ungar}, {and}
  \bibinfo{person}{David~M Pennock}.} \bibinfo{year}{2002}\natexlab{}.
\newblock \showarticletitle{Methods and metrics for cold-start
  recommendations}. In \bibinfo{booktitle}{\emph{Proceedings of the 25th annual
  international ACM SIGIR conference on Research and development in information
  retrieval}}. \bibinfo{pages}{253--260}.
\newblock


\bibitem[\protect\citeauthoryear{Simonyan and Zisserman}{Simonyan and
  Zisserman}{2014}]%
        {simonyan2014very}
\bibfield{author}{\bibinfo{person}{Karen Simonyan} {and}
  \bibinfo{person}{Andrew Zisserman}.} \bibinfo{year}{2014}\natexlab{}.
\newblock \showarticletitle{Very deep convolutional networks for large-scale
  image recognition}.
\newblock \bibinfo{journal}{\emph{arXiv preprint arXiv:1409.1556}}
  (\bibinfo{year}{2014}).
\newblock


\bibitem[\protect\citeauthoryear{Sohl-Dickstein, Weiss, Maheswaranathan, and
  Ganguli}{Sohl-Dickstein et~al\mbox{.}}{2015}]%
        {sohl2015deep}
\bibfield{author}{\bibinfo{person}{Jascha Sohl-Dickstein},
  \bibinfo{person}{Eric Weiss}, \bibinfo{person}{Niru Maheswaranathan}, {and}
  \bibinfo{person}{Surya Ganguli}.} \bibinfo{year}{2015}\natexlab{}.
\newblock \showarticletitle{Deep unsupervised learning using nonequilibrium
  thermodynamics}. In \bibinfo{booktitle}{\emph{International conference on
  machine learning}}. PMLR, \bibinfo{pages}{2256--2265}.
\newblock


\bibitem[\protect\citeauthoryear{Song, Kim, Nowozin, Ermon, and Kushman}{Song
  et~al\mbox{.}}{2017}]%
        {song2017pixeldefend}
\bibfield{author}{\bibinfo{person}{Yang Song}, \bibinfo{person}{Taesup Kim},
  \bibinfo{person}{Sebastian Nowozin}, \bibinfo{person}{Stefano Ermon}, {and}
  \bibinfo{person}{Nate Kushman}.} \bibinfo{year}{2017}\natexlab{}.
\newblock \showarticletitle{Pixeldefend: Leveraging generative models to
  understand and defend against adversarial examples}.
\newblock \bibinfo{journal}{\emph{arXiv preprint arXiv:1710.10766}}
  (\bibinfo{year}{2017}).
\newblock


\bibitem[\protect\citeauthoryear{Tang, Du, He, Yuan, Tian, and Chua}{Tang
  et~al\mbox{.}}{2019}]%
        {tang2019adversarial}
\bibfield{author}{\bibinfo{person}{Jinhui Tang}, \bibinfo{person}{Xiaoyu Du},
  \bibinfo{person}{Xiangnan He}, \bibinfo{person}{Fajie Yuan},
  \bibinfo{person}{Qi Tian}, {and} \bibinfo{person}{Tat-Seng Chua}.}
  \bibinfo{year}{2019}\natexlab{}.
\newblock \showarticletitle{Adversarial training towards robust multimedia
  recommender system}.
\newblock \bibinfo{journal}{\emph{IEEE Transactions on Knowledge and Data
  Engineering}} \bibinfo{volume}{32}, \bibinfo{number}{5}
  (\bibinfo{year}{2019}), \bibinfo{pages}{855--867}.
\newblock


\bibitem[\protect\citeauthoryear{Veit, Kovacs, Bell, McAuley, Bala, and
  Belongie}{Veit et~al\mbox{.}}{2015}]%
        {veit2015learning}
\bibfield{author}{\bibinfo{person}{Andreas Veit}, \bibinfo{person}{Balazs
  Kovacs}, \bibinfo{person}{Sean Bell}, \bibinfo{person}{Julian McAuley},
  \bibinfo{person}{Kavita Bala}, {and} \bibinfo{person}{Serge Belongie}.}
  \bibinfo{year}{2015}\natexlab{}.
\newblock \showarticletitle{Learning visual clothing style with heterogeneous
  dyadic co-occurrences}. In \bibinfo{booktitle}{\emph{Proceedings of the IEEE
  international conference on computer vision}}. \bibinfo{pages}{4642--4650}.
\newblock


\bibitem[\protect\citeauthoryear{Wang, Xu, Feng, Lin, He, and Chua}{Wang
  et~al\mbox{.}}{2023}]%
        {wang2023diffusion}
\bibfield{author}{\bibinfo{person}{Wenjie Wang}, \bibinfo{person}{Yiyan Xu},
  \bibinfo{person}{Fuli Feng}, \bibinfo{person}{Xinyu Lin},
  \bibinfo{person}{Xiangnan He}, {and} \bibinfo{person}{Tat-Seng Chua}.}
  \bibinfo{year}{2023}\natexlab{}.
\newblock \showarticletitle{Diffusion Recommender Model}.
\newblock \bibinfo{journal}{\emph{arXiv preprint arXiv:2304.04971}}
  (\bibinfo{year}{2023}).
\newblock


\bibitem[\protect\citeauthoryear{Xia, Huang, Xu, Zhao, Yin, and Huang}{Xia
  et~al\mbox{.}}{2022}]%
        {xia2022hypergraph}
\bibfield{author}{\bibinfo{person}{Lianghao Xia}, \bibinfo{person}{Chao Huang},
  \bibinfo{person}{Yong Xu}, \bibinfo{person}{Jiashu Zhao},
  \bibinfo{person}{Dawei Yin}, {and} \bibinfo{person}{Jimmy Huang}.}
  \bibinfo{year}{2022}\natexlab{}.
\newblock \showarticletitle{Hypergraph contrastive collaborative filtering}. In
  \bibinfo{booktitle}{\emph{Proceedings of the 45th International ACM SIGIR
  conference on research and development in information retrieval}}.
  \bibinfo{pages}{70--79}.
\newblock


\bibitem[\protect\citeauthoryear{Xu, Evans, and Qi}{Xu et~al\mbox{.}}{2017}]%
        {xu2017feature}
\bibfield{author}{\bibinfo{person}{Weilin Xu}, \bibinfo{person}{David Evans},
  {and} \bibinfo{person}{Yanjun Qi}.} \bibinfo{year}{2017}\natexlab{}.
\newblock \showarticletitle{Feature squeezing: Detecting adversarial examples
  in deep neural networks}.
\newblock \bibinfo{journal}{\emph{arXiv preprint arXiv:1704.01155}}
  (\bibinfo{year}{2017}).
\newblock


\bibitem[\protect\citeauthoryear{Yang, Zhang, Song, Hong, Xu, Zhao, Zhang, Cui,
  and Yang}{Yang et~al\mbox{.}}{2023}]%
        {yang2023diffusion}
\bibfield{author}{\bibinfo{person}{Ling Yang}, \bibinfo{person}{Zhilong Zhang},
  \bibinfo{person}{Yang Song}, \bibinfo{person}{Shenda Hong},
  \bibinfo{person}{Runsheng Xu}, \bibinfo{person}{Yue Zhao},
  \bibinfo{person}{Wentao Zhang}, \bibinfo{person}{Bin Cui}, {and}
  \bibinfo{person}{Ming-Hsuan Yang}.} \bibinfo{year}{2023}\natexlab{}.
\newblock \showarticletitle{Diffusion models: A comprehensive survey of methods
  and applications}.
\newblock \bibinfo{journal}{\emph{Comput. Surveys}} \bibinfo{volume}{56},
  \bibinfo{number}{4} (\bibinfo{year}{2023}), \bibinfo{pages}{1--39}.
\newblock


\bibitem[\protect\citeauthoryear{Yin, Cui, Huang, Wang, Wu, and Zhou}{Yin
  et~al\mbox{.}}{2015}]%
        {yin2015joint}
\bibfield{author}{\bibinfo{person}{Hongzhi Yin}, \bibinfo{person}{Bin Cui},
  \bibinfo{person}{Zi Huang}, \bibinfo{person}{Weiqing Wang},
  \bibinfo{person}{Xian Wu}, {and} \bibinfo{person}{Xiaofang Zhou}.}
  \bibinfo{year}{2015}\natexlab{}.
\newblock \showarticletitle{Joint modeling of users' interests and mobility
  patterns for point-of-interest recommendation}. In
  \bibinfo{booktitle}{\emph{Proceedings of the 23rd ACM international
  conference on Multimedia}}. \bibinfo{pages}{819--822}.
\newblock


\bibitem[\protect\citeauthoryear{Yin, Cui, Sun, Hu, and Chen}{Yin
  et~al\mbox{.}}{2014}]%
        {yin2014lcars}
\bibfield{author}{\bibinfo{person}{Hongzhi Yin}, \bibinfo{person}{Bin Cui},
  \bibinfo{person}{Yizhou Sun}, \bibinfo{person}{Zhiting Hu}, {and}
  \bibinfo{person}{Ling Chen}.} \bibinfo{year}{2014}\natexlab{}.
\newblock \showarticletitle{LCARS: A spatial item recommender system}.
\newblock \bibinfo{journal}{\emph{ACM Transactions on Information Systems
  (TOIS)}} \bibinfo{volume}{32}, \bibinfo{number}{3} (\bibinfo{year}{2014}),
  \bibinfo{pages}{1--37}.
\newblock


\bibitem[\protect\citeauthoryear{Yin, Qu, Chen, Yuan, Zheng, Long, Xia, Shi,
  and Zhang}{Yin et~al\mbox{.}}{2024}]%
        {yin2024device}
\bibfield{author}{\bibinfo{person}{Hongzhi Yin}, \bibinfo{person}{Liang Qu},
  \bibinfo{person}{Tong Chen}, \bibinfo{person}{Wei Yuan},
  \bibinfo{person}{Ruiqi Zheng}, \bibinfo{person}{Jing Long},
  \bibinfo{person}{Xin Xia}, \bibinfo{person}{Yuhui Shi}, {and}
  \bibinfo{person}{Chengqi Zhang}.} \bibinfo{year}{2024}\natexlab{}.
\newblock \showarticletitle{On-Device Recommender Systems: A Comprehensive
  Survey}.
\newblock \bibinfo{journal}{\emph{arXiv preprint arXiv:2401.11441}}
  (\bibinfo{year}{2024}).
\newblock


\bibitem[\protect\citeauthoryear{Yin, Liu, Gong, and Li}{Yin
  et~al\mbox{.}}{2023}]%
        {yin2023securing}
\bibfield{author}{\bibinfo{person}{Minglei Yin}, \bibinfo{person}{Bin Liu},
  \bibinfo{person}{Neil~Zhenqiang Gong}, {and} \bibinfo{person}{Xin Li}.}
  \bibinfo{year}{2023}\natexlab{}.
\newblock \showarticletitle{Securing Visually-Aware Recommender Systems: An
  Adversarial Image Reconstruction and Detection Framework}.
\newblock \bibinfo{journal}{\emph{arXiv preprint arXiv:2306.07992}}
  (\bibinfo{year}{2023}).
\newblock


\bibitem[\protect\citeauthoryear{Yuan, Nguyen, He, Chen, and Yin}{Yuan
  et~al\mbox{.}}{2023a}]%
        {yuan2023manipulating}
\bibfield{author}{\bibinfo{person}{Wei Yuan}, \bibinfo{person}{Quoc Viet~Hung
  Nguyen}, \bibinfo{person}{Tieke He}, \bibinfo{person}{Liang Chen}, {and}
  \bibinfo{person}{Hongzhi Yin}.} \bibinfo{year}{2023}\natexlab{a}.
\newblock \showarticletitle{Manipulating Federated Recommender Systems:
  Poisoning with Synthetic Users and Its Countermeasures}.
\newblock \bibinfo{journal}{\emph{arXiv preprint arXiv:2304.03054}}
  (\bibinfo{year}{2023}).
\newblock


\bibitem[\protect\citeauthoryear{Yuan, Yang, Qu, Ye, Nguyen, and Yin}{Yuan
  et~al\mbox{.}}{2024}]%
        {yuan2024robust}
\bibfield{author}{\bibinfo{person}{Wei Yuan}, \bibinfo{person}{Chaoqun Yang},
  \bibinfo{person}{Liang Qu}, \bibinfo{person}{Guanhua Ye},
  \bibinfo{person}{Quoc Viet~Hung Nguyen}, {and} \bibinfo{person}{Hongzhi
  Yin}.} \bibinfo{year}{2024}\natexlab{}.
\newblock \showarticletitle{Robust Federated Contrastive Recommender System
  against Model Poisoning Attack}.
\newblock \bibinfo{journal}{\emph{arXiv preprint arXiv:2403.20107}}
  (\bibinfo{year}{2024}).
\newblock


\bibitem[\protect\citeauthoryear{Yuan, Yuan, Yang, Nguyen, and Yin}{Yuan
  et~al\mbox{.}}{2023b}]%
        {yuan2023manipulating1}
\bibfield{author}{\bibinfo{person}{Wei Yuan}, \bibinfo{person}{Shilong Yuan},
  \bibinfo{person}{Chaoqun Yang}, \bibinfo{person}{Quoc Viet~Hung Nguyen},
  {and} \bibinfo{person}{Hongzhi Yin}.} \bibinfo{year}{2023}\natexlab{b}.
\newblock \showarticletitle{Manipulating Visually-aware Federated Recommender
  Systems and Its Countermeasures}.
\newblock \bibinfo{journal}{\emph{ACM Transactions on Information Systems}}
  (\bibinfo{year}{2023}).
\newblock


\bibitem[\protect\citeauthoryear{Zeng, Liu, Wang, Qiu, Xie, Tai, Tang, and
  Yuille}{Zeng et~al\mbox{.}}{2019}]%
        {zeng2019adversarial}
\bibfield{author}{\bibinfo{person}{Xiaohui Zeng}, \bibinfo{person}{Chenxi Liu},
  \bibinfo{person}{Yu-Siang Wang}, \bibinfo{person}{Weichao Qiu},
  \bibinfo{person}{Lingxi Xie}, \bibinfo{person}{Yu-Wing Tai},
  \bibinfo{person}{Chi-Keung Tang}, {and} \bibinfo{person}{Alan~L Yuille}.}
  \bibinfo{year}{2019}\natexlab{}.
\newblock \showarticletitle{Adversarial attacks beyond the image space}. In
  \bibinfo{booktitle}{\emph{Proceedings of the IEEE/CVF Conference on Computer
  Vision and Pattern Recognition}}. \bibinfo{pages}{4302--4311}.
\newblock


\bibitem[\protect\citeauthoryear{Zhang, Gao, and Rao}{Zhang
  et~al\mbox{.}}{2021a}]%
        {zhang2021defense}
\bibfield{author}{\bibinfo{person}{Shudong Zhang}, \bibinfo{person}{Haichang
  Gao}, {and} \bibinfo{person}{Qingxun Rao}.} \bibinfo{year}{2021}\natexlab{a}.
\newblock \showarticletitle{Defense against adversarial attacks by
  reconstructing images}.
\newblock \bibinfo{journal}{\emph{IEEE Transactions on Image Processing}}
  \bibinfo{volume}{30} (\bibinfo{year}{2021}), \bibinfo{pages}{6117--6129}.
\newblock


\bibitem[\protect\citeauthoryear{Zhang, Yin, Chen, Huang, Cui, and Zhang}{Zhang
  et~al\mbox{.}}{2021b}]%
        {zhang2021graph}
\bibfield{author}{\bibinfo{person}{Shijie Zhang}, \bibinfo{person}{Hongzhi
  Yin}, \bibinfo{person}{Tong Chen}, \bibinfo{person}{Zi Huang},
  \bibinfo{person}{Lizhen Cui}, {and} \bibinfo{person}{Xiangliang Zhang}.}
  \bibinfo{year}{2021}\natexlab{b}.
\newblock \showarticletitle{Graph embedding for recommendation against
  attribute inference attacks}. In \bibinfo{booktitle}{\emph{Proceedings of the
  Web Conference 2021}}. \bibinfo{pages}{3002--3014}.
\newblock


\bibitem[\protect\citeauthoryear{Zhang, Yin, Chen, Huang, Nguyen, and
  Cui}{Zhang et~al\mbox{.}}{2022}]%
        {zhang2022pipattack}
\bibfield{author}{\bibinfo{person}{Shijie Zhang}, \bibinfo{person}{Hongzhi
  Yin}, \bibinfo{person}{Tong Chen}, \bibinfo{person}{Zi Huang},
  \bibinfo{person}{Quoc Viet~Hung Nguyen}, {and} \bibinfo{person}{Lizhen Cui}.}
  \bibinfo{year}{2022}\natexlab{}.
\newblock \showarticletitle{Pipattack: Poisoning federated recommender systems
  for manipulating item promotion}. In \bibinfo{booktitle}{\emph{Proceedings of
  the Fifteenth ACM International Conference on Web Search and Data Mining}}.
  \bibinfo{pages}{1415--1423}.
\newblock


\bibitem[\protect\citeauthoryear{Zhao, Lin, Feng, Wang, and Wen}{Zhao
  et~al\mbox{.}}{2022}]%
        {zhao2022revisiting}
\bibfield{author}{\bibinfo{person}{Wayne~Xin Zhao}, \bibinfo{person}{Zihan
  Lin}, \bibinfo{person}{Zhichao Feng}, \bibinfo{person}{Pengfei Wang}, {and}
  \bibinfo{person}{Ji-Rong Wen}.} \bibinfo{year}{2022}\natexlab{}.
\newblock \showarticletitle{A revisiting study of appropriate offline
  evaluation for top-N recommendation algorithms}.
\newblock \bibinfo{journal}{\emph{ACM Transactions on Information Systems}}
  \bibinfo{volume}{41}, \bibinfo{number}{2} (\bibinfo{year}{2022}),
  \bibinfo{pages}{1--41}.
\newblock


\bibitem[\protect\citeauthoryear{Zheng, Qu, Chen, Zheng, Shi, and Yin}{Zheng
  et~al\mbox{.}}{2024}]%
        {zheng2024poisoning}
\bibfield{author}{\bibinfo{person}{Ruiqi Zheng}, \bibinfo{person}{Liang Qu},
  \bibinfo{person}{Tong Chen}, \bibinfo{person}{Kai Zheng},
  \bibinfo{person}{Yuhui Shi}, {and} \bibinfo{person}{Hongzhi Yin}.}
  \bibinfo{year}{2024}\natexlab{}.
\newblock \showarticletitle{Poisoning Decentralized Collaborative Recommender
  System and Its Countermeasures}.
\newblock \bibinfo{journal}{\emph{arXiv preprint arXiv:2404.01177}}
  (\bibinfo{year}{2024}).
\newblock


\bibitem[\protect\citeauthoryear{Zheng, Qu, Cui, Shi, and Yin}{Zheng
  et~al\mbox{.}}{2023}]%
        {zheng2023automl}
\bibfield{author}{\bibinfo{person}{Ruiqi Zheng}, \bibinfo{person}{Liang Qu},
  \bibinfo{person}{Bin Cui}, \bibinfo{person}{Yuhui Shi}, {and}
  \bibinfo{person}{Hongzhi Yin}.} \bibinfo{year}{2023}\natexlab{}.
\newblock \showarticletitle{Automl for deep recommender systems: A survey}.
\newblock \bibinfo{journal}{\emph{ACM Transactions on Information Systems}}
  \bibinfo{volume}{41}, \bibinfo{number}{4} (\bibinfo{year}{2023}),
  \bibinfo{pages}{1--38}.
\newblock


\end{thebibliography}










\end{document}